# High-fidelity 3D multi-slab diffusion MRI using Slab-shifting for Harmonized 3D Acquisition and Reconstruction with Profile Encoding Networks (SHARPEN)

Ziyu Li[1], Karla L. Miller[1*], Wenchuan Wu[1*]

[1]Oxford Centre for Integrative Neuroimaging, FMRIB, Nuffield Department of Clinical Neurosciences, University of Oxford, Oxford, UK.

* indicates equal contributions.

Correspondence to: Ziyu Li, Ph.D., Oxford Centre for Integrative Neuroimaging (OxCIN), FMRIB, John Radcliffe Hospital, Oxford, OX3 9DU, UK. E-mail: ziyu.li@ndcn.ox.ac.uk.

**Abstract**

Three-dimensional (3D) multi-slab imaging is a promising approach for high-resolution in-vivo diffusion MRI (dMRI) due to its compatibility with short TR=1-2 s that provides optimal signal-to-noise ratio (SNR) efficiency. A major challenge, however, is slab boundary artifacts arising from non-ideal slab-selective RF excitation. Specifically, the non-rectangular slab profile leads to reduced signal intensity at slab boundaries. In addition, slab profiles can extend into adjacent slabs and introduce inter-slab crosstalk, where overlapping excitation shortens the local TR and limits T1 recovery. To mitigate slab boundary artifacts without increasing scan time, we build on the slab profile encoding concept and propose Slab-shifting for Harmonized 3D Acquisition and Reconstruction with Profile Encoding Networks (SHARPEN). For different diffusion directions, SHARPEN uses variable inter-volume field-of-view shifts along the slice direction to provide complementary slab profile encoding without prolonging acquisition. Slab profiles are estimated with a lightweight self-supervised neural network that exploits intrinsic consistency across shifted acquisitions and known physical properties of slab profiles and diffusion images. The corrected images are reconstructed using the estimated profiles. SHARPEN was validated using simulated and prospectively acquired high-resolution in-vivo data and demonstrates accurate slab profile estimation and boundary artifacts correction, even in the presence of inter-volume motion. Importantly, SHARPEN does not require high-quality reference training data and supports subject-specific training. Its efficient GPU-based implementation delivers substantially faster and more accurate correction than the existing NPEN method, yielding quantitative profiles along the slice direction that closely match those from reference 2D acquisitions. SHARPEN enables high-quality dMRI at 0.7 mm isotropic resolution on a 3T clinical scanner, highlighting its potential to advance high-fidelity submillimeter dMRI for neuroscience research.

## 1. Introduction

Diffusion MRI (dMRI) provides unique, noninvasive insights into the microstructure and connectivity of the living human brain. Recent technical advances have focused on improving image fidelity and spatial resolution [1, 2], enabling more detailed characterization of microscopic tissue features and complex fiber architectures that are critical for neuroscience and clinical research. A key constraint for achieving high-resolution dMRI is signal-to-noise ratio (SNR), as diffusion encoding intrinsically attenuates signal and SNR decreases as voxel size is reduced.

To improve SNR efficiency (i.e., SNR per unit scan time), 3D multi-slab dMRI has been proposed. In this approach, the imaging volume is divided into multiple slabs, each typically consisting of 10-20 slices, which are excited and spatially encoded as 3D k-space. Within each slab, 3D echo-planar imaging (EPI) is typically employed with the slice direction (kz) defined by phase encoding, enabling the reconstruction of thin, sharp slices. Because the number of slabs is much smaller than the total number of slices for large brain coverages, 3D multi-slab imaging is compatible with short TR=1-2 s, the regime in which spin-echo-based dMRI achieves optimal SNR efficiency [3-5]. Using this framework, we recently demonstrated in vivo 3D dMRI at 0.53 mm isotropic resolution, among the highest resolutions achieved so far on 3T clinical scanners [6].

Despite these advantages, the image quality of 3D multi-slab dMRI is still limited by slab boundary artifacts. These artifacts arise because radiofrequency (RF) excitation profiles are not perfectly rectangular. As a result, signal intensity is intrinsically reduced at slab boundaries. Additionally, excitation from one slab can leak into adjacent slabs (i.e., slab crosstalk). Consequently, slices in overlapped regions are repeatedly excited with an effectively shorter TR, causing incomplete T1 recovery and further signal loss due to saturation effects [7]. Moreover, slab profiles exceeding the kz encoding field-of-view (FOV) lead to slice aliasing. Collectively, these effects manifest as slab boundary artifacts (typically observed as low-signal banding along the slice direction) that can compromise the accuracy of downstream diffusion analyses.

Several strategies have been proposed to mitigate slab boundary artifacts. Slice aliasing can be addressed by oversampling along kz to expand the FOV and discarding outermost boundary slices where aliasing is most severe [3, 7, 8]. However, addressing the reduced signal at slab boundaries is more challenging. A common approach is to increase the number of overlapping slices between slabs to improve SNR through averaging measurements across slabs. Unfortunately, this strategy exacerbates saturation effects and leads to residual artifacts, particularly at short TRs [3, 7]. Even with additional post-processing techniques such as weighted-averaging with 1D slab profile and Fourier band-pass filtering, slab boundary artifacts remain apparent at short TRs (e.g., ≤2 s) [7].

Slab-shifted acquisition has been proposed as an alternative solution [9-11]. In this approach, the acquisition of each volume is repeated with several FOV shifts along the slab direction, such that the final volume can be reconstructed using only central slices (unaffected by slab boundary artifacts) from each shifted acquisition while discarding the affected slices at slab boundaries. Slab oversampling is employed in each acquisition to prevent slice aliasing, thereby enabling complete slice coverage while avoiding slab boundary artifacts. Although these strategies can effectively minimize slab boundary artifacts, they require substantially increased scan times. Previous implementations doubled [9, 10] or even tripled [11] the acquisition time to achieve high-quality 3D multi-slab dMRI. Additionally, these approaches are inherently sensitive to inter-scan motion, which can introduce inconsistencies between the shifted acquisitions.

To reduce slab boundary artifacts without substantially increasing scan time, slab profile encoding methods have been introduced [12-14]. The original profile encoding (PEN) framework [12] modeled multi-slab acquisition as a linear encoding problem that can be solved via linear inversion, assuming accurate knowledge of the slab excitation profiles. In practice, however, accurate profile estimation is difficult, particularly at short TRs where strong saturation effects cannot be captured by the calibration scans used in PEN.

To address this limitation, nonlinear profile encoding (NPEN) was subsequently proposed [13]. NPEN reformulates the problem as a nonlinear optimization that jointly estimates the slab profile and the corrected image, exploiting prior knowledge such as in-plane smoothness of the slab profile and periodicity of slab boundary artifacts. This joint optimization enables improved profile estimation and artifact correction, and has been shown to perform well even at TR=2 s [13]. Nevertheless, important challenges remain. First, jointly estimating both the slab profile and the image makes NPEN a highly ill-posed problem. In challenging scenarios, such as low SNR, strong saturation effects, or severe B0/B1+ inhomogeneity, NPEN may still fail to recover accurate profiles and artifact-free images. Indeed, in our prior work on submillimeter in vivo dMRI where SNR was extremely limited [6], slab boundary artifacts remained noticeable after NPEN correction. Second, NPEN relies on iterative Gauss-Newton optimization, which is computationally expensive and leads to long runtimes.

More recently, deep learning approaches have been explored to overcome these limitations. CPEN [14] introduced unrolled convolutional neural networks (CNNs) to solve the nonlinear profile encoding problem, achieving faster reconstruction via forward inference of trained networks. However, CPEN relies on supervised training with high-quality, slab-artifact–free reference data. Such training data typically require long acquisition times (17.2 min for two volumes at 1.3 mm [14]), making them motion-sensitive and impractical for many studies, while the generalizability of trained models across scanners and studies remains uncertain.

Motivated by these challenges, we propose SHARPEN: Slab-shifting for Harmonized 3D Acquisition and Reconstruction with Profile Encoding Networks. SHARPEN integrates slab-shifted acquisition with a self-supervised deep learning framework to estimate slab profiles and correct for slab boundary artifacts. Instead of acquiring multiple slab shifts for each diffusion volume as used in previous slab-shifted acquisition methods, SHARPEN applies inter-volume FOV shifts across different diffusion directions, providing complementary encoding information without increasing total scan time. A lightweight, self-supervised CNN then estimates the slab profiles by enforcing profile consistency across shifted acquisitions, leveraging the similarity of direction-averaged diffusion contrast, and uses the estimated profiles to reconstruct the corrected images. We validate SHARPEN using both realistic simulations and prospectively acquired data, including a high-resolution 0.7 mm isotropic dataset. We demonstrate that SHARPEN can be trained using single-subject data without any external training datasets and achieve superior artifact suppression and substantially faster reconstruction compared to NPEN. These results highlight the potential of SHARPEN to enable high-fidelity submillimeter 3D multi-slab dMRI for neuroscience and clinical applications.

## 2. Theory

### *2.1. Review of slab profile encoding in 3D multi-slab imaging*

In 3D multi-slab imaging, slab-selective RF pulses are used to excite multiple slabs. Within each slab, thin slices are resolved using 3D Fourier encoding along the slice direction. The original PEN framework [12] models this process as a linear encoding problem:

$$S\rho = d, \tag{1}$$

where $\rho$ denotes the desired artifact-free 3D image volume, $d$ represents the acquired multi-slab k-space data, and $S$ is the slab profile encoding operator. The operator $S$ accounts for signal modulation due to slab excitation profiles and saturation effects from slab crosstalk, as well as slice aliasing. If $S$ is known accurately, $\rho$ could be recovered via a linear least-squares inversion:

$$\rho = S^{-1}d = (S^H S)^{-1} S^H d, \tag{2}$$

where $S^{-1}$ denotes the least-squares (pseudo-inverse) of the slab profile encoding operator $S$, and $S^H$ is the Hermitian transpose of $S$. In practice, however, accurate estimation of $S$ is challenging. The original PEN estimates slab profiles using Bloch simulations or short calibration scans, neither of which fully capture saturation effects or profile distortions caused by B0/B1+ inhomogeneities. Inaccurate profile estimation consequently degrades the quality of the reconstructed image $\rho$.

To more accurately model the profile, NPEN [13] treats both $S$ and $\rho$ as unknowns, defining $x = [\rho, S]^T$ and reformulating the forward model as a nonlinear operator:

$$E(x) = d, \tag{3}$$

where $E$ maps the image and profile to the acquired data.

NPEN solves Eq. 3 using an iteratively regularized Gauss-Newton algorithm. At iteration $n$, the update $\Delta x_n$ is obtained by solving:

$$\min_{\Delta x_n} \left|\left|E'(x_n)\Delta x_n - \left(d - E(x_n)\right)\right|\right|_2^2 + \alpha \left|\left|x_n + \Delta x_n - x_0\right|\right|_2^2 + \beta \left|\left|W_u F \rho_n\right|\right|_2^2, \tag{4}$$

where $E'(x_n)$ is the Fréchet derivative of $E$ at the current estimate $x_n$, $x_0$ is the initial guess, and the update is applied as $x_{n+1} = x_n + \Delta x_n$. The second term is a Tikhonov regularization that stabilizes the inversion, while the third term penalizes residual slab boundary artifacts. Here, $F$ is Fourier transform and $W_u$ is a weighting matrix that emphasizes frequencies corresponding to the known periodicity of slab boundary artifacts. To enforce in-plane smoothness of the slab profile $S$, NPEN reparametrizes the unknowns as $x = [\rho, W_S F S]^T$, where $W_S$ penalizes high spatial frequency components of $S$ in each kz plane.

Despite these regularizations, the joint estimation problem in Eq. 3 remains ill-posed. Accurate recovery of $\rho$ and $S$ is particularly challenging in the presence of strong saturation effects (introducing tissue-dependent profile variations), pronounced B0/B1+ inhomogeneity (violating the assumed periodicity of slab boundary artifacts), or low SNR (further degrading conditioning). The initial slab profile estimate in NPEN is generated from a 1D slab profile estimation representing the global signal variations across a slab but not accounting for in-plane spatial variations. The ill-posed nature also makes NPEN sensitive to the inaccuracy of the initial slab profile estimate. The Gauss-Newton optimization in Eq. 4 is also computationally expensive, especially for high-resolution datasets and when using a large number of iterations for better estimates.

To further improve robustness and computational efficiency, CPEN [14] introduced a deep learning-based approach to iteratively solve the non-linear problem in Eq. 3. Specifically, CPEN replaces each Gauss-Newton update with a convolutional neural network (CNN) $g_{a_n}$, parametrized by $a_n$:

$$x_{n+1} = g_{a_n}(x_n; \rho_r). \tag{5}$$

where $\rho_r$ is an artifact-free reference image used to guide the reconstruction. The CNN parameters are learned by minimizing the loss function:

$$\mathcal{L}_{\text{CPEN}}(x_n) = \left|\left|E(x_n) - d\right|\right|_2^2 + \eta_1 l_{\text{SSIM}}(\rho_n, \rho_r)$$
$$+\eta_2\left|\left|W_u F \rho_n\right|\right|_2^2 + \eta_3 \text{Var}(S_n) + \eta_4\left|\left|x_n - x_0\right|\right|_2^2, \tag{6}$$

where $l_{\text{SSIM}}$ denotes the structural similarity (SSIM) loss (lower values correspond to higher similarity), $\left|\left|W_u F \rho_n\right|\right|_2^2$ suppresses the periodic artifacts in the same way as NPEN, $\text{Var}(S_n)$ promotes in-plane smoothness of the slab profile by minimizing its variance, and the final term is a Tikhonov regularization. Conceptually, Eq. 6 closely mirrors the NPEN objective in Eq. 4, with the key distinction being the use of the artifact-free reference image $\rho_r$. In CPEN, $\rho_r$ is obtained using slab-shifted acquisitions with extensive slice oversampling to ensure slab-boundary artifacts are minimized in the combined image, which requires approximately threefold longer scan times (17.2 min for two volumes at 1.3 mm). The supervised model achieved robust performance and fast inference once trained. However, the need for such high-quality reference data substantially limits the practical applicability of CPEN due to the prolonged scan time and increased motion sensitivity. Moreover, training $N$ (number of iterations) unrolled CNNs incurs considerable time and computational cost, further limiting scalability and practical deployment.

### *2.2. SHARPEN: Slab-shifting for Harmonized 3D Acquisition and Reconstruction with Profile Encoding Networks*

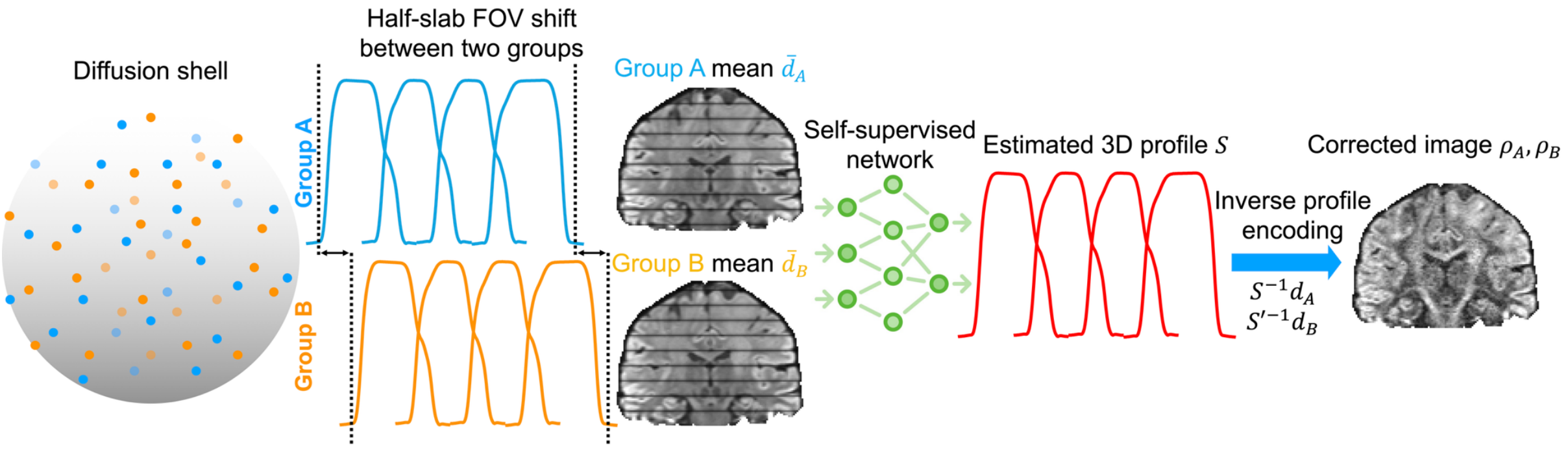

**Figure 1. Slab-shifting for Harmonized 3D Acquisition and Reconstruction with Profile Encoding Networks (SHARPEN) framework.** Within each diffusion shell, diffusion directions are divided into two groups (A and B). The data are acquired with a field-of-view (FOV) shift along the slice direction between the two groups. Group-wise averaged images $\bar{d}_A$and $\bar{d}_B$provide complementary slab-profile-encoding information and are used to train a self-supervised network to estimate the 3D slab profile $S$. The estimated profile and its shifted counterpart ($S'$) are then applied to all volumes within the shell to reconstruct the corrected images $\rho_A$ and $\rho_B$.

The proposed SHARPEN framework is illustrated in Fig. 1. Within each b-value shell, diffusion directions are divided into two groups, denoted as A and B. Data from the two groups are acquired with a relative half-a-slab FOV shift along the slice direction. This inter-volume shift ensures that slab-boundary slices in one group correspond to slab-center slices in the other, thereby providing complementary slab-encoding information. Importantly, this strategy does not increase the total scan time, as each volume is acquired only once.

For reconstruction, we train a 3D CNN to reconstruct the 3D slab profile $S$ from the averaged images of these two groups:

$$S = f_\theta(\bar{d}_A, \bar{d}_B), \tag{7}$$

where $f_\theta$ is the CNN parameterized by $\theta$, $\bar{d}_A, \bar{d}_B$ are group-averaged images from group A and B, respectively. The network is trained in a self-supervised manner without requiring artifact-free reference images by exploiting the consistency between the shifted acquisitions and known structural properties of slab profiles and corrected images. Specifically, the training minimizes:

$$\mathcal{L}_{\text{SHARPEN}}(S) = \mathcal{L}_{\text{shift}}(S) + \lambda_{\text{profile}}\mathcal{L}_{\text{profile}}(S) + \lambda_{\text{slice}}\mathcal{L}_{\text{slice}}(S) + \lambda_{\text{smooth}}\mathcal{L}_{\text{smooth}}(S), \tag{8}$$

where the weights $\lambda_*$ balance each term's contribution.

The loss $\mathcal{L}_{\text{shift}}$ exploits the similarity of corrected images and the slab profiles from the two shifted acquisitions (i.e., groups A and B). Although groups A and B consist of diffusion-weighted images (DWIs) acquired along different directions, their within-group averaged, corrected images share similar contrast corresponding to the trace of the diffusion tensor, i.e., $\bar{\rho}_A \approx \bar{\rho}_B$, provided that each group contains a sufficient number of diffusion directions. Moreover, the slab excitation profile is assumed to be invariant to the shift apart from a spatial displacement, such that knowledge of the profile $S$ directly determines its shifted counterpart $S'$. This assumption is supported by the fact that tissue composition as well as B0/B1+ field variations over the half-slab shift distance are small. Accordingly, $\mathcal{L}_{\text{shift}}$ penalizes the discrepancy between the corrected averaged images:

$$\mathcal{L}_{\text{shift}}(S) = \left|\left|\bar{\rho}_A - \bar{\rho}_B\right|\right|_2^2 = \left|\left|S^{-1}\bar{d}_A - S'^{-1}\bar{d}_B\right|\right|_2^2, \tag{9}$$

where $S'$ denotes a spatially shifted version of the slab profile $S$.

The loss $\mathcal{L}_{\text{profile}}$ encourages physically plausible slab profiles. In particular, the slab profile is expected to have an amplitude close to unity at slab-center locations. For each slab, two subsets of slice locations are identified: a center set $\mathcal{Z}_{\text{center}}$, corresponding to the central slices of that slab, and a boundary set $\mathcal{Z}_{\text{boundary}}$, corresponding to the $n$ outermost slices at the slab boundaries ($n$=2 in this study). Owing to the half-slab shift between the two acquisitions, boundary slices in one group spatially align with center slices in the other. Under the assumptions that the direction-averaged corrected images satisfy $\bar{\rho}_A \approx \bar{\rho}_B$ and that the profile amplitude at center locations satisfies $S_{\text{center}} \approx 1$, the ratio of the averaged measurements provides an estimate of the boundary profile: $\bar{d}_B./\bar{d}_A = S_{\text{boundary}}\bar{\rho}_B./S_{\text{center}}\bar{\rho}_A \approx S_{\text{boundary}}$. Therefore,

$$\mathcal{L}_{\text{profile}}(S) = \sum_{z\in\mathcal{Z}_{\text{center}}} ||S_z - 1||_2^2 + \sum_{z\in\mathcal{Z}_{\text{boundary}}} \left|\left|S_z - \left(\bar{d}_B \,./\bar{d}_A\,\right)_z\right|\right|_2^2. \quad (10)$$

The loss $\mathcal{L}_{\text{slice}}$ exploits the expectation that, after slab boundary correction, in-plane-averaged slice-wise signal intensity should vary smoothly along the slice direction. This constraint is applied only to slab-boundary slices to avoid unnecessary regularization of slab-center slices, which are minimally affected by slab boundary artifacts:

$$\mathcal{L}_{\text{slice}}(S) = \sum_{z\in\mathcal{Z}_{\text{boundary}}} |\text{mean}(\bar{\rho}_A)_{z+1} - \text{mean}(\bar{\rho}_A)_z| = \sum_{z\in\mathcal{Z}_{\text{boundary}}} \left|\text{mean}\left(S^{-1}\bar{d}_A\right)_{z+1} - \text{mean}\left(S^{-1}\bar{d}_A\right)_z\right|, \quad (11)$$

where mean(·) denotes in-plane averaging over the spatial dimensions (x,y), applied prior to computing finite differences along z.

The loss $\mathcal{L}_{\text{smooth}}$ enforces in-plane (x-y) smoothness of the slab profile to improve the conditioning of the profile estimation problem. Similar constraints have been used in NPEN [13] and CPEN [14]. Here, the smoothness is imposed by minimizing the in-plane total variation of $S$:

$$\mathcal{L}_{smooth}(S) = \sum_{x,y,z} \left(\left|S_{x+1,y,z} - S_{x,y,z}\right| + \left|S_{x,y+1,z} - S_{x,y,z}\right|\right). \quad (12)$$

The slab profile is estimated once and then applied to all volumes within the shell to obtain the corrected images:

$$\rho_A = S^{-1}d_A, \qquad \rho_B = S'^{-1}d_B. \quad (13)$$

As established in PEN [12], once an accurate slab profile is available, it can be applied consistently across volumes provided subject motion is not severe. This reflects the fact that B0, B1+, and tissue relaxation properties (which dictate saturation effects) typically vary smoothly in space, meaning that moderate inter-volume motion does not substantially alter the slab excitation profiles.

## 3. Methods

### *3.1. Simulation Data*

Preprocessed diffusion data from 10 subjects from Human Connectome Project (HCP) [15, 16] were used to simulate 3D multi-slab data with slab boundary artifacts, and validate the efficacy of SHARPEN. In this study, 8 b=0 and 90 b=1000 s/mm$^2$ volumes at 1.25 mm isotropic resolution were used.

Ten slab excitation profiles were simulated, each consisting of 10 effective slices with 20% oversampling (12 slices per slab) and a 1-slice overlap between adjacent slabs, encoding 92 slices in total (FOV$_z$=115 mm). Slab profiles were generated by simulating slab-selective RF excitation using windowed sinc pulses, with the pulse bandwidth chosen to achieve a target slab full-width-at-half-maximum (FWHM) of 9 mm. Spatial B0 inhomogeneity was modeled using slab-dependent off-resonance effects, reflecting stronger field variations toward inferior brain regions, while B1+ inhomogeneity was incorporated as smooth slice-wise amplitude Gaussian modulation along the slab direction.

To model tissue-dependent saturation effects in overlapped slices, voxel-wise T1 maps were simulated from white matter, gray matter, and cerebrospinal fluid (CSF) partial-volume estimates obtained using SynthSeg [17] with T1w images as inputs (T1 relaxation times: white matter/gray matter/CSF=0.85/1.30/4.00 s [18]). A saturation factor $R_{sat}$, representing the relative signal attenuation due to incomplete longitudinal recovery, was computed assuming TR = 2 s and applied multiplicatively to overlapped slices to model repeated excitation:

$$R_{sat} = \frac{1 - e^{-\frac{TR}{T1}}}{1 - e^{-\frac{TR}{2T1}}} . \tag{14}$$

A mild 3D Gaussian smoothing (kernel size σ = (3,3,0.6) voxels) was finally applied to the slab profiles to reflect their inherent spatial smoothness.

To simulate the proposed slab-shifted acquisition, the slab profile was additionally shifted by half a slab (5 slices), and non-shifted and shifted profiles were applied in an interleaved manner across diffusion directions from HCP. Conventional multi-slab acquisition was simulated by applying only the non-shifted slab profile to all volumes. Finally, Rician noise was added to the profile-modulated multi-slab data, assuming SNR$_{DWI}$=10.

### *3.2. Prospectively Acquired Data*

In-plane single-shot [5] and segmented [6] 3D multi-slab spin-echo dMRI sequences with 2D navigators were adapted to enable slab shifting along the slice direction by introducing half-slab frequency offsets in the excitation and refocusing slab-selective RF pulses. The sequences were implemented using the open-source, scanner agnostic framework “Pulseq” [19] to enable broader accessibility and applications.

In vivo experiments were performed on a 3T MRI system (Siemens Prisma, Erlangen, Germany) using a 32-channel receive head coil. Written informed consent was obtained from all participants in accordance with local institutional ethics approval. Diffusion data were acquired using this slab-shifted 3D multi-slab sequence at 1.1 mm isotropic resolution from four subjects without in-plane segmentation to evaluate the efficacy of SHARPEN, and at 0.7 mm isotropic resolution from one subject with in-plane segmentation to demonstrate SHARPEN's capability for high-fidelity submillimeter dMRI. Key acquisition parameters for both protocols are summarized in Table 1.

| Res. ($mm^3$) | Matrix size | $N_{\text{slice}}$ per slab /OS/overlap[a] | TE1/TE2/TR (ms)[b] | $N_{\text{shot}}$/R[c] | $ES_{eff}$ (ms)[d] | $T_{\text{acq}}$ per vol. | #b0/ DWI[e] | $T_{\text{acq}}$ |
|---|---|---|---|---|---|---|---|---|
| 1.1 | 220×220×110 | 10/20% /1 | 65/133/2200 | 1/3 | 0.27 | 26.4 s | 4/32 | 16 min |
| 0.7 | 300×300×170 | 15/13.3%/1 | 76/187/2700 | 2/4 | 0.28 | 91.8 s | 4/32 | 61 min |

**Table 1. Key parameters for prospective dMRI acquisitions.** a. $N_{\text{slice}}$ denotes the number of slices. Oversampling along kz (OS) is applied to reduce slice aliasing (e.g., 10 slices/slab with 20% oversampling leads to 12 acquired slices). b. TE1 is the imaging echo time and TE2 is the navigator echo time. c. $N_{\text{shot}}$ is the number of acquired in-plane shots and R is the in-plane acceleration factor of each shot. For the 0.7 mm protocol, the 2 acquired ky segments are evenly spaced. d. $ES_{eff}$ refers to the effective echo spacing (i.e., echo spacing/R). e. DWIs are acquired at b=1000 s/$mm^2$. b=0 data are acquired with opposite phase encoding directions to enable distortion correction (2 anterior-posterior (AP) and 2 posterior-anterior (PA)). The 0.7 mm b=0 data are acquired without acceleration to ensure high image quality (i.e., 4 shots acquired).

Both protocols covered approximately 120 mm along the slice direction. Partial Fourier (PF = 3/4) was applied in the phase-encoding direction, and slab excitation was achieved using Shinnar–Le Roux (SLR) RF pulses [20]. For the 0.7 mm protocol, the refocusing flip angle was reduced to 160° to reduce slab boundary saturation effects [6]. A whole-brain gradient echo coil calibration scan (~2 min acquisition time) was performed to enable parallel imaging reconstruction.

To enable slab-shifted acquisition, the 32 diffusion directions were divided into two groups. The directions were partitioned such that each group of 16 directions maintained approximately uniform coverage of the sphere, resulting in similar trace-weighted contrast in the group-wise mean diffusion-weighted images. One group was acquired using slab-shifted acquisition and the other using non-shifted acquisition, and the two groups were interleaved throughout the scan to ensure that both groups were similarly affected by motion, thereby reducing potential bias when estimating slab profiles using the group-mean images.

For the 1.1 mm protocol, to enable direct comparison with conventional non-shifted multi-slab acquisitions, both slab-shifted and non-shifted acquisitions were applied to each diffusion direction, resulting in a total scan time of 32 min. To assess SHARPEN's robustness to inter-volume motion between shifted acquisitions, an additional shifted b=0 volume was acquired at the end of the session in one subject. This volume was expected to exhibit relatively large subject motion compared to the non-shifted b=0 volume acquired at the beginning of the session. To quantitatively evaluate the accuracy of slab boundary correction, an additional 2D reference dataset was acquired for each subject with matched slice thickness but lower in-plane resolution. This reference dataset was collected using the Siemens product 2D simultaneous multi-slice (SMS) diffusion-weighted EPI sequence with the following parameters: matrix size=136×136×110, resolution=1.5×1.5×1.1 $mm^3$, R=2, PF=3/4, multi-band factor=2, TE/TR=65/6000 ms, matched diffusion directions with the 1.1 mm data, and a total acquisition time of approximately 5 min. The 2D reference dataset was reconstructed using the scanner's standard online reconstruction.

For 3D multi-slab dMRI reconstruction, all 32-coil k-space data were compressed to 8 coils [21]. The 1.1 mm data were reconstructed using SPIRiT with motion-induced phase error correction [22], which is minimally regularized to isolate and assess slab boundary artifact correction:

$$x = \arg\min_{x} \sum_{i}^{N_{kz}} \left|\left|D_i F P_i F^{-1} x - y_i\right|\right|_2^2 + \lambda_1 \left|\left|(G - I)x\right|\right|_2^2, \tag{15}$$

where $x$ is the desired fully-sampled multi-coil k-space data to be reconstructed, $N_{kz}$ is the total number of kz planes, $D_i$ is the shot-sampling mask for the $i^{th}$ shot, $F$ is the Fourier transform, $P_i$ is the motion-induced phase term for the $i^{th}$ shot, $y_i$ is the acquired data for the $i^{th}$ shot, $G$ is the SPIRiT kernel trained on coil calibration data, with a kernel size of 5×5, and $I$ is the identity matrix. Motion-induced phase maps $P_i$ were estimated from navigator data acquired for each shot. Specifically, 2D navigator images were reconstructed using GRAPPA [23], followed by k-space Hamming filtering (window size 32×32) to suppress noise, exploiting the expected spatial smoothness of the phase variations [5]. The resulting phase images were extracted and incorporated into the reconstruction. $\lambda_1$ was empirically set to 10. Eq. 15 was solved using a conjugate gradient algorithm.

The submillimeter 0.7 mm data were reconstructed using DnSPIRiT that iteratively alternates image denoising with forward-model-based reconstruction [6], reflecting the substantially lower SNR at submillimeter resolution and the need for additional regularization to enable high-fidelity reconstruction. At iteration $n$:

$$x_n = \arg\min_{x_n} \sum_{i}^{N_{\text{shot}} \times N_{kz}} \left|\left|D_i F P_i F^{-1} x_n - y_i\right|\right|_2^2 + \lambda_1 \left|\left|(G - I)x_n\right|\right|_2^2 + \lambda_2 \left|\left|F^{-1} x_n - z_{n-1}\right|\right|_2^2,$$
$$z_n = \Phi(F^{-1} x_n), \tag{16}$$

where $\Phi$ is the denoiser. DnSPIRiT has been shown to effectively suppress noise for submillimeter dMRI reconstruction while minimizing denoising-induced bias and image blurring by enforcing data consistency [6]. Similar to the previous work [6], BM4D [24] using adaptive noise level estimation with Rician noise distribution was used as the denoiser. $\lambda_1$ and $\lambda_2$ were empirically set to 10 and 1, and Eq. 16 was run for 5 iterations.

### *3.3. Slab Correction using SHARPEN*

The outermost oversampled slices on both sides of each slab, which predominantly contain noise and slice aliasing artifacts, were discarded prior to slab boundary correction. This step effectively mitigated slice aliasing and simplified the profile encoding model to focus on slab excitation and saturation effects.

A lightweight 3D CNN was used as $f_\theta$ to estimate the slab profile (Supplementary Fig. 1). The network takes as input two slab-shifted, slice position-aligned images concatenated along the channel dimension and outputs an estimated slab profile. To reduce GPU memory requirements, the input slab images were processed in a block-wise manner, with blocks of size $N_b \times N_b \times N_{\text{slice}}$. The block size was set to $N_b = 64$ for HCP simulation and the 1.1 mm data, and $N_b = 96$ for the 0.7 mm data. The CNN consists of five consecutive 3D convolutional layers with 3×3×3 kernels and 32 feature channels, each followed by a ReLU activation. A final 3×3×3 convolution with a single output channel is then applied, followed by a sigmoid function that maps the final slab profile to (0, 1). The network contains approximately 0.11 million trainable parameters and can be trained using only a single pair of slab-shifted images from one subject.

The network was implemented in PyTorch (https://pytorch.org/) and trained in a subject-specific, self-supervised manner by minimizing Eq. 8, with each subject trained independently using only that subject's data. Hyperparameters were tuned via grid search using HCP simulations, where the objective was to minimize the difference between the estimated slab profiles and the ground-truth profiles, yielding $\lambda_{\text{profile}}$=2, $\lambda_{\text{slice}}$=3, and $\lambda_{\text{smooth}}$=5. Training employed the AdamW optimizer (weight decay=$10^{-4}$) [25] with a ReduceLROnPlateau learning-rate scheduler (reduction factor 0.5, patience 5) [26]. Each model was trained for 200 epochs on the b=0 shell. On the b=1000 s/mm$^2$, the model pretrained on b=0 shell was fine-tuned for 50 epochs using slab-shifted mean DWIs rather than being trained from scratch to accelerate the convergence. The initial learning rates were empirically set to $5 \times 10^{-5}$, $1 \times 10^{-4}$, and $4 \times 10^{-5}$ for the HCP, 1.1 mm, and 0.7 mm datasets, respectively, and $4 \times 10^{-6}, 4 \times 10^{-5}$, and $1 \times 10^{-5}$ for b=1000 s/mm$^2$ fine-tuning. Training loss curves for the b=0 shell from one representative subject in each dataset are shown in Supplementary Fig. 2, demonstrating stable and robust convergence across datasets. All experiments were performed on an NVIDIA A30 GPU, with per-epoch training times of approximately 1.3 s, 1.9 s, and 5.5 s for the HCP, 1.1 mm, and 0.7 mm datasets, respectively, resulting in total per-subject processing times (training, inference, plus slab profile correction to all volumes) of approximately 6 min, 8 min, and 23 min.

For comparison, conventional slab boundary correction strategies, including overlap averaging and NPEN, were applied to non-shifted acquisitions. The overlap-average method averaged overlapped slices after discarding the oversampled slices. NPEN followed the original implementation [13], with hyperparameters tuned separately for each dataset. For the HCP data, NPEN was applied to both conventional non-shifted acquisitions and the proposed slab-shifted acquisitions (Fig. 1), the latter denoted as NPEN-S. NPEN converged rapidly on the HCP data but required substantially more iterations for the prospective 1.1 mm and 0.7 mm datasets. The corresponding per-volume processing times were approximately 8 min, 85 min, and 266 min for the HCP (98 volumes), 1.1 mm (36 volumes), and 0.7 mm (36 volumes) datasets, respectively. The per-subject processing times will be similar to per-volume processing times when high-performance parallel computing is available.

To evaluate SHARPEN's robustness to inter-volume motion, slab boundary correction was performed on the 1.1 mm dataset using both motion-free and motion-corrupted shifted b=0 volumes. Two correction strategies were compared. First, SHARPEN-based correction was applied separately using each shifted volume, with one network trained using motion-free data and another trained using motion-corrupted data. Second, a conventional slab-shifted combination strategy [11] was evaluated as a baseline, as it performs well in the absence of motion by discarding artifact-contaminated boundary slices via 'cut-and-combine' of slab-center slices but is inherently sensitive to inter-scan motion. Results obtained using motion-free and motion-corrupted shifted volumes were then compared, enabling assessment of SHARPEN's robustness to inter-volume motion relative to the conventional combination approach.

### *3.4. Image Analyses*

Image post-processing was performed using the FMRIB Software Library (FSL) [27]. For each subject's prospectively acquired data following slab correction, a whole-brain field map was estimated from blip-reversed b=0 volumes using "topup" [28]. This field map, together with all diffusion-weighted data, was then input to "eddy" [29] to correct for susceptibility-induced distortions, eddy current effects, and subject motion. The 2D reference data were rigidly aligned to the 1.1 mm data using "flirt". Diffusion tensor imaging (DTI) model fitting was performed on all diffusion datasets using "dtifit".

For the HCP data, the corrected b=0 and b=1000 s/mm$^2$ images, as well as DTI-derived metrics including mean diffusivity (MD), fractional anisotropy (FA), and diffusion tensor estimation, were compared with the corresponding ground-truth results. Within brain masks, image similarity was quantified using the normalized root mean squared error (NRMSE). The accuracy of MD and FA estimates was assessed using mean absolute error (MAE), while diffusion tensor estimation accuracy was quantified by the Frobenius norm of the difference between the estimated and ground-truth tensor matrices. For the prospectively acquired 1.1 mm data, slab boundary correction performance was evaluated by comparing slice profiles from the corrected 3D multi-slab data with those from the 2D reference dataset. To mitigate the impact of contrast differences arising from different TE/TR, slice profiles were compared using MD maps, which provide contrast normalization across acquisitions. Pearson correlation coefficients (r) were computed between the slice profiles of the corrected 3D multi-slab data, obtained using overlap averaging, NPEN, and SHARPEN, and the corresponding 2D reference profiles to quantify profile similarity.

## 4. Results

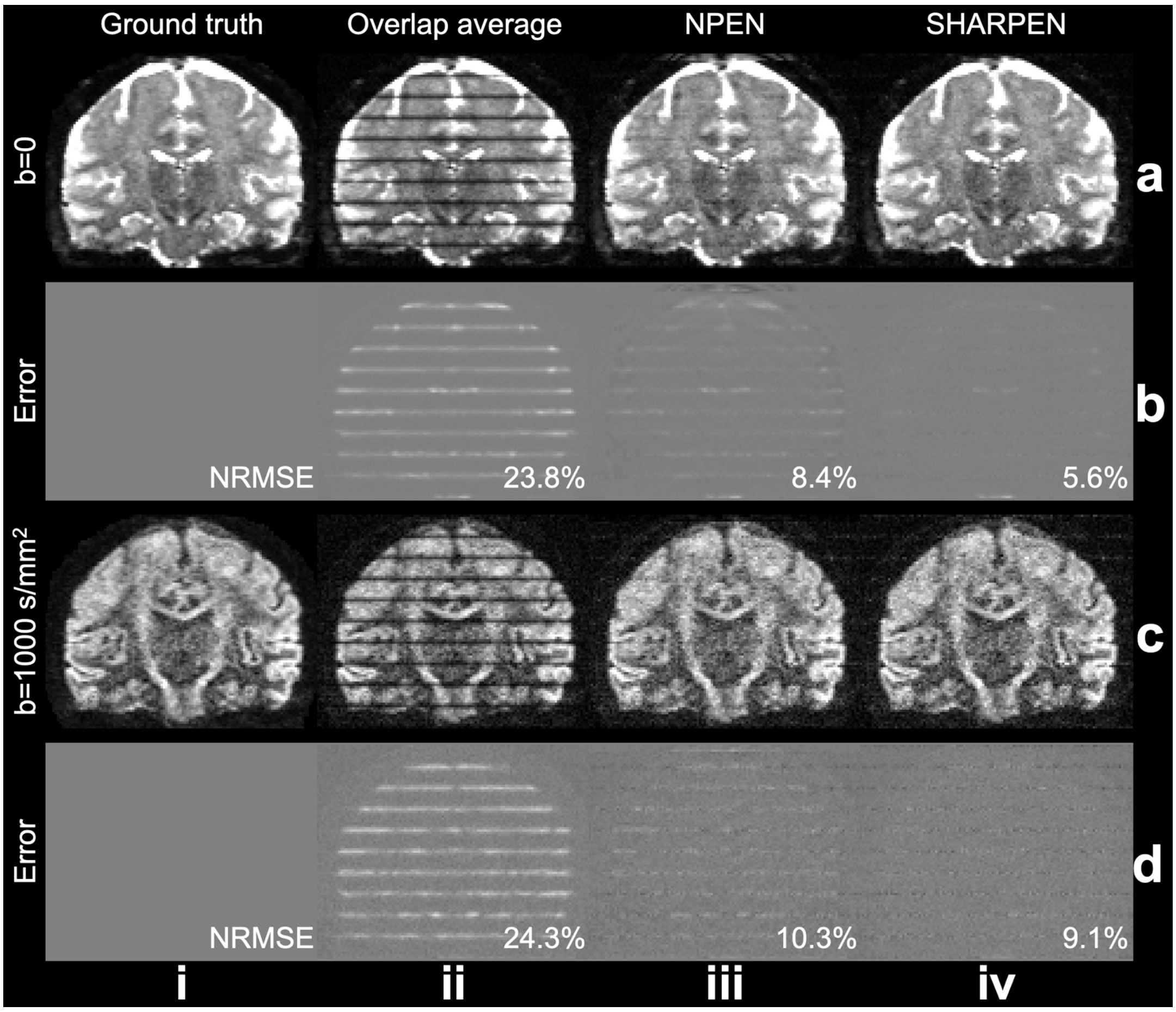

**Figure 2. Single-volume image correction results on simulated HCP data.** Representative coronal slices from b=0 (a) and b=1000 s/mm$^2$ (c) simulated HCP data are shown for ground truth (i), overlap average (ii, i.e., averaging of overlapping slab slices without explicit profile correction), NPEN (iii), and SHARPEN (iv) results. Corresponding residual maps (b, d) highlight differences from the ground truth, with normalized root mean squared errors (NRMSE) quantifying each method's image similarity with the ground truth.

Slab boundary correction results for representative single volumes from the simulated HCP dataset are shown in Fig. 2. Pronounced slab boundary artifacts are evident in the overlap average reconstruction, which averages overlapping slab slices without explicit profile correction (Fig. 2, ii). Both NPEN and SHARPEN substantially reduce these artifacts. However, SHARPEN yields visibly fewer residual artifacts, particularly in the b=0 images (Fig. 2a, b, iii vs. iv). For the b=1000 s/mm$^2$ data, the residual map from SHARPEN predominantly reflects added noise, whereas NPEN residuals still contain structured slab boundary artifacts (Fig. 2d, iii vs. iv).

Figure 3 further demonstrates SHARPEN's ability to accurately estimate slab excitation profiles. The 1D slab profiles (slice-wise averages of the 3D profiles) estimated by SHARPEN show closer agreement with the simulated ground truth than those obtained with NPEN (Fig. 3a, i). Notably, NPEN estimates slab-center profile values exceeding unity. This behavior primarily arises from the NPEN initialization, which is derived from a 1D signal profile along the slice direction that captures not only local slab-profile modulation but also global signal variation. As a result, the initialized profile and consequently the optimized profile can absorb global intensity variations, leading to slab-center values greater than one and a more uniform signal along the slice direction after correction. In contrast, SHARPEN explicitly constrains the profile estimation through cross-shift consistency, yielding profiles that more closely reflect the underlying excitation characteristics.

Furthermore, SHARPEN recovers spatially varying saturation effects at slab boundaries in-plane, which are not captured by NPEN (Fig. 3a,b,iii vs. iv). These differences are most evident in CSF regions of the b = 0 images, where strong saturation effects are present. They are less pronounced in the b = 1000 s/mm$^2$ images, because diffusion weighting substantially attenuates CSF signal, leaving little residual signal to be corrected. This interpretation is consistent with Fig. 2, where NPEN exhibits greater difficulty correcting b = 0 data than b = 1000 s/mm² data, whereas SHARPEN provides robust correction in both cases.

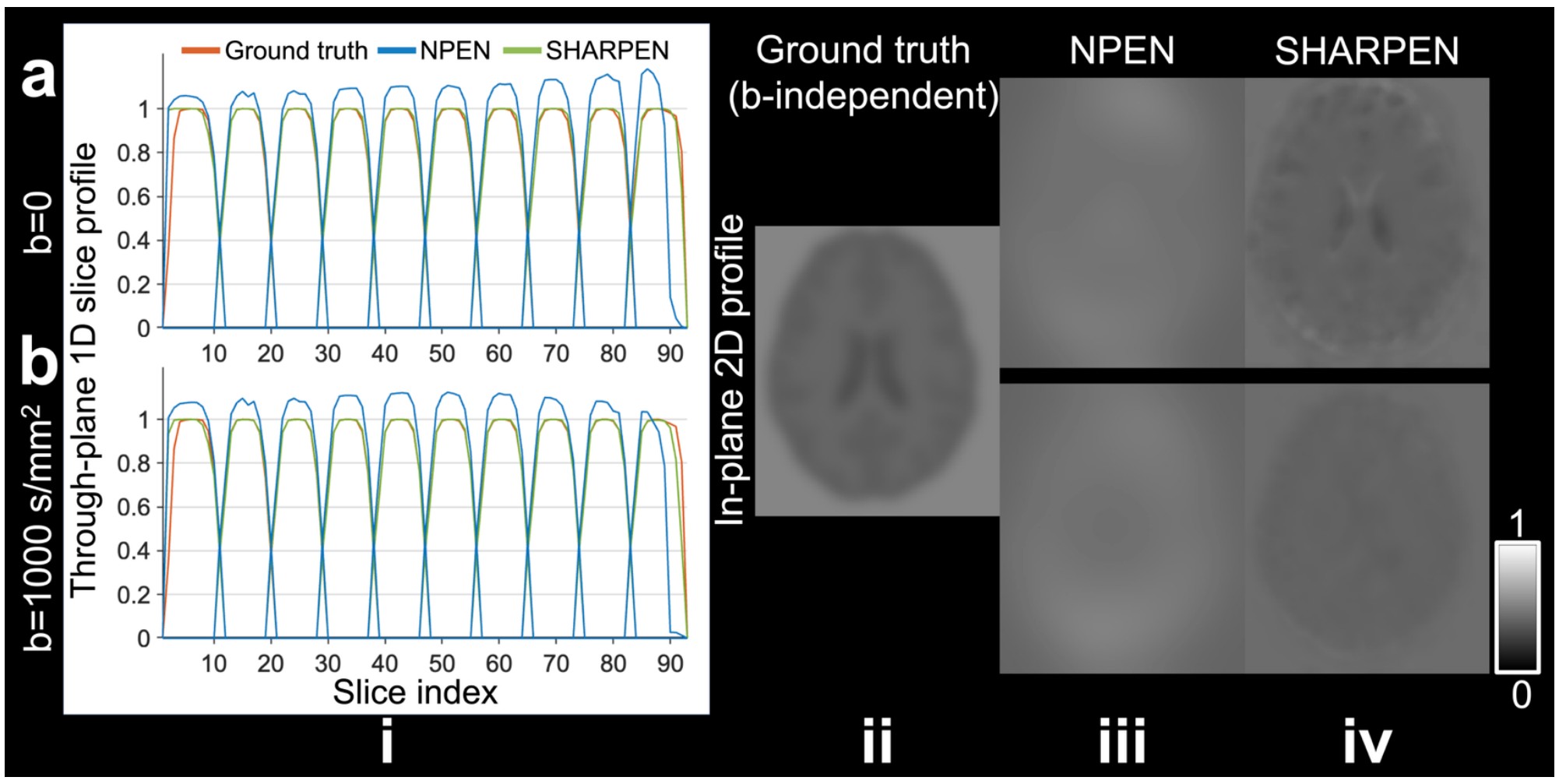

**Figure 3. Slab profile estimation on simulated HCP data.** The estimated 1D slab profiles along the slice direction (i) and in-plane 2D profiles for NPEN (iii) and SHARPEN (iv) are compared against the simulation ground truth (independent of b-value, ii) for b=0 (a) and b=1000s/mm$^2$ (b) data from a representative HCP subject.

The effective slab boundary correction achieved by SHARPEN translates into improved diffusion tensor imaging (DTI) fitting accuracy on the simulated HCP data (Fig. 4). Two NPEN-based correction strategies are shown for comparison: NPEN applied to conventional non-shifted data (Fig. 4, ii) and NPEN applied to SHARPEN-style (Fig. 1) data with interdigitated shifting (NPEN-S; Fig. 4, iii). Both NPEN and NPEN-S exhibit residual slab-related errors in the derived MD and FA maps (Fig. 4, ii, iii), which likely arises from NPEN's artifact suppression regularization term ($\beta||W_u F\rho_n||_2^2$ in Eq. 4), which acts as a band-pass filter along the slice direction and would introduce additional artifacts if the periodicity assumption of artifacts is not met. While slab shifting reduces MD errors in NPEN-S relative to NPEN, FA errors are increased, likely due to the higher variance introduced by slab shifting. In contrast, SHARPEN substantially suppresses slab boundary artifacts and yields lower quantitative errors in both MD and FA, with residual maps dominated by added random noise rather than structured artifacts (Fig. 4, iv). These results demonstrate that SHARPEN's performance gains arise not only from the slab-shifted acquisition strategy, but also from the improved slab boundary correction enabled by the proposed network.

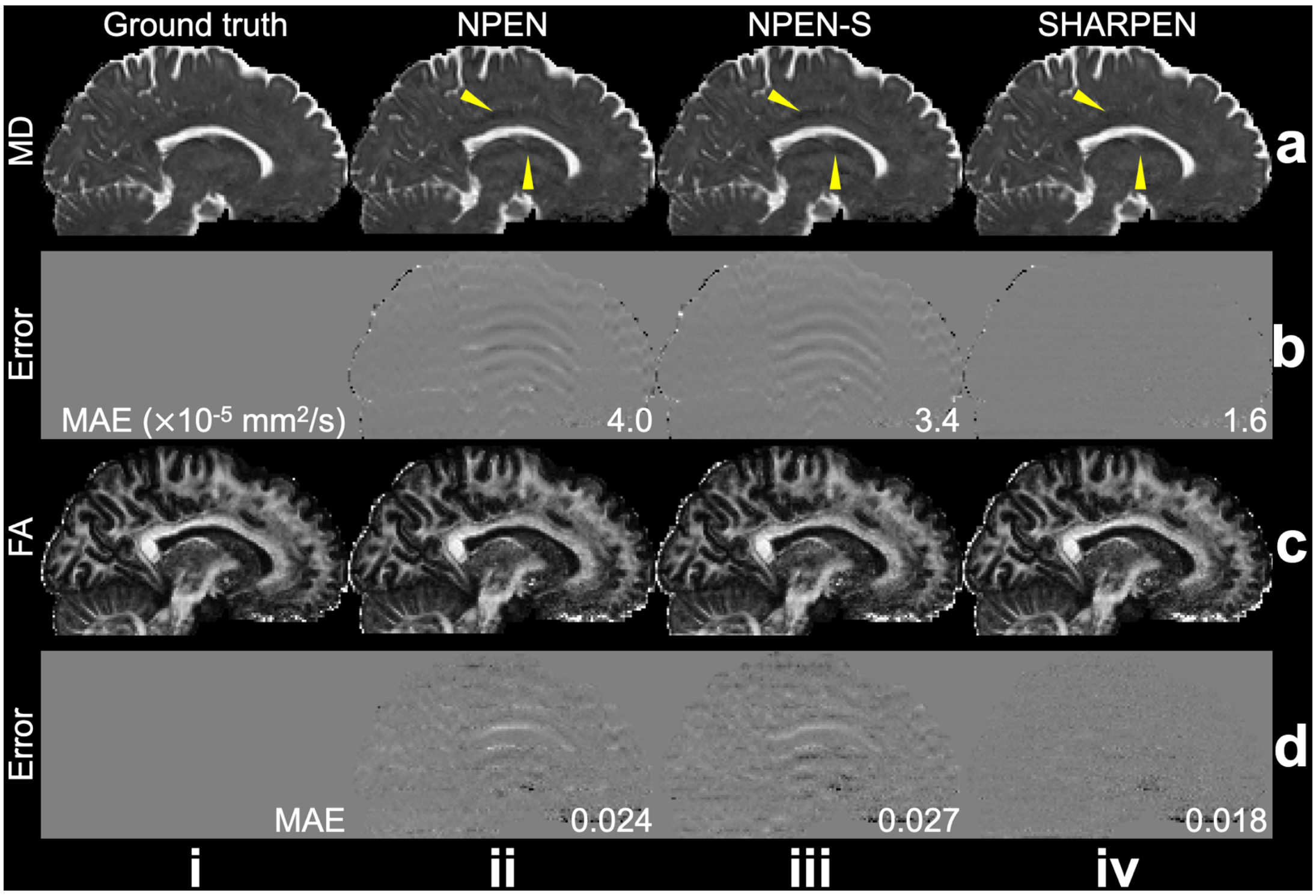

**Figure 4. DTI fitting results on simulated HCP data.** Representative sagittal slices of fractional anisotropy (FA) (a) and mean diffusivity (MD) (c) maps derived from simulated HCP data are shown for different correction strategies: ground truth (i), NPEN with conventional non-shifted acquisition (ii), NPEN with shifted acquisition (NPEN-S, iii), and SHARPEN (iv). Corresponding residual maps (b, d) show differences from the ground truth, with mean absolute errors (MAE) quantifying fitting accuracy. Yellow arrows mark regions of SHARPEN's improvement over NPEN.

| | Frobenius norm of tensor error (×$10^{-5}$) | MD error (×$10^{-5}$ $mm^2/s$) | FA error (×$10^{-2}$) |
|---|---|---|---|
| NPEN | 11.76±0.93 | 3.87±0.50 | 2.32±0.08 |
| NPEN-S | 11.48±0.87 | 3.26±0.46 | 2.79±0.06 |
| SHARPEN | **7.48±0.42** | **1.63±0.16** | **1.89±0.11** |

**Table 2. Group-level DTI fitting accuracy comparison on simulated HCP data.** Group-level (mean±std) DTI fitting errors for simulated data from 10 HCP subjects, including the Frobenius norm of the tensor estimation error and mean absolute errors of mean diffusivity (MD) and fractional anisotropy (FA), are compared across NPEN with conventional non-shifted acquisition, NPEN with SHARPEN-style slab-shifted acquisition (NPEN-S), and SHARPEN. For each metric, the best performance is highlighted in **bold**.

Quantitative comparisons of group-level DTI fitting accuracy for NPEN, NPEN-S, and SHARPEN across 10 HCP subjects are summarized in Table 2, with slice-wise error profiles shown in Fig. 5. Consistent with the qualitative observations in Fig. 4, NPEN-S improves mean MD accuracy relative to NPEN by 15.8%, but at the expense of degraded FA accuracy by 20.3%, resulting in only marginal improvement in overall diffusion tensor estimation accuracy by 2.4%. SHARPEN substantially improves DTI fitting accuracy across all evaluated metrics. Compared with NPEN, SHARPEN reduces the MD error by 57.9%, the FA error by 18.5%, and the tensor estimation error by 36.4% across 10 HCP subjects. NPEN exhibits pronounced MD errors at high slice indices, where slow-varying biases in the estimated slab profiles most evident in the b=0 data lead to systematic mis-estimation of diffusion metrics. Importantly, these reductions are consistently observed across all subjects. In addition, the slab-shifted acquisition strategy also reduces the variance of slice-wise errors for SHARPEN relative to conventional NPEN, as boundary slices benefit from having half of the diffusion directions sampled at slab centers, leading to more uniform accuracy across slices (Fig. 5).

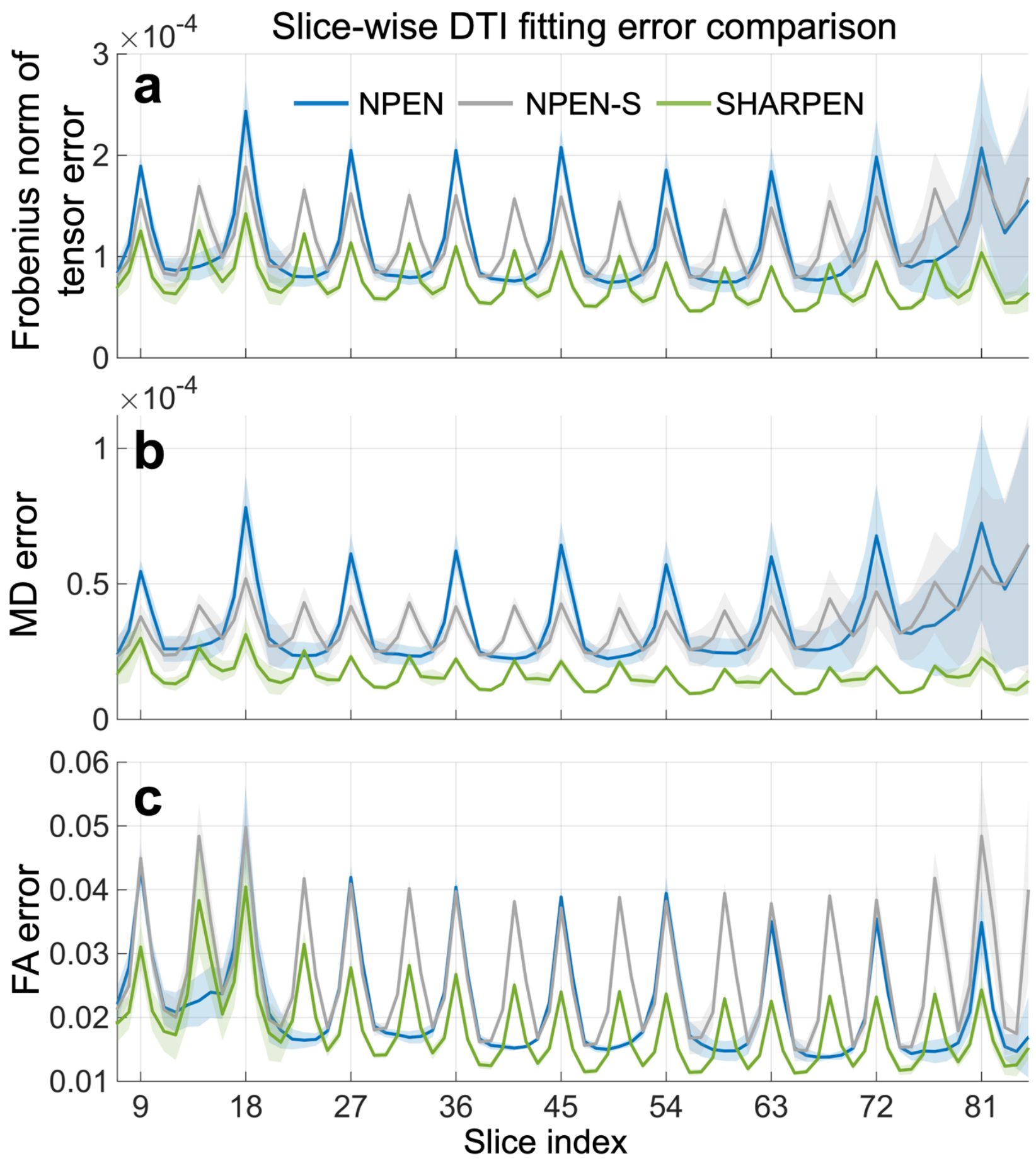

**Figure 5. Group-level slice-wise DTI fitting accuracy on simulated HCP data.** Slice-wise DTI fitting errors for simulated data from 10 HCP subjects, including the Frobenius norm of the tensor estimation error (a) and mean absolute errors of mean diffusivity (MD, b) and fractional anisotropy (FA, c), are compared across NPEN with conventional non-shifted acquisition, NPEN with slab-shifted acquisition (NPEN-S), and SHARPEN. Curves represent the mean error across subjects, and shaded regions indicate ±1 standard deviation. Only slices containing valid brain tissue across all subjects are displayed.

The prospectively acquired 1.1 mm data further validate the efficacy of SHARPEN. Single-volume 3D multi-slab images exhibit high SNR and sharp anatomical detail (Fig. 6). However, overlap-averaging reconstruction suffers from pronounced slab boundary artifacts that substantially degrade image quality (Fig. 6a, b, i). Although NPEN reduces these artifacts, residual slab boundary effects remain evident, particularly in the b=0 images (Fig. 6a, ii). These residual artifacts are likely exacerbated by the short TR (2.2 s) and the sharp SLR excitation pulse, which together induce strong signal decrease due to saturation and steep profile roll-off at slab boundaries. In contrast, SHARPEN suppresses the artifacts to produce clean images for both b=0 and b=1000 s/mm$^2$ data (Fig. 6a, b, iii). Examination of slab boundary slices further highlights these differences: NPEN results show pronounced saturation artifacts in b=0 images and amplified noise in b=1000 s/mm$^2$ images, particularly near the ventricles (Fig. 6c, d, i).

SHARPEN effectively mitigates both effects, yielding boundary-slice images that are highly similar to the shifted reference, which are slab-center slices from a slab-shifted reference acquisition where this same slice is at the slab center (Fig. 6c, d, ii vs. iii). Mean b=0 and mean diffusion-weighted images show consistent trends (Supplementary Fig. 3).

DTI fitting results for the 1.1 mm data (Fig. 7) mirror these observations. The 3D multi-slab acquisition enables high-SNR, high-resolution DTI maps within a 16 min scan. Residual slab boundary artifacts in the overlap average and NPEN-corrected images propagate into the DTI results (Fig. 7, i, ii), whereas these artifacts are largely suppressed by SHARPEN (Fig. 7, iii), yielding cross-slice variations comparable to those of the 2D reference dataset (Fig. 7, iv). Notably, slab boundary artifacts are generally less prominent in DTI maps than in raw images, particularly for overlap averaging, as shared excitation profiles between b=0 and DWI lead to partial normalization during tensor fitting. The remaining artifacts might arise from inter-volume motion, which alters relative image positioning, and partial volume effects, where signal contributions are modulated by both relaxation and diffusion properties [13]. It is also worth noting that the 3D multi-slab data have higher spatial resolution (1.1 mm isotropic) than the 2D reference dataset ($1.5 \times 1.5 \times 1.1$ mm$^3$), enabling improved visualization of fine anatomical details.

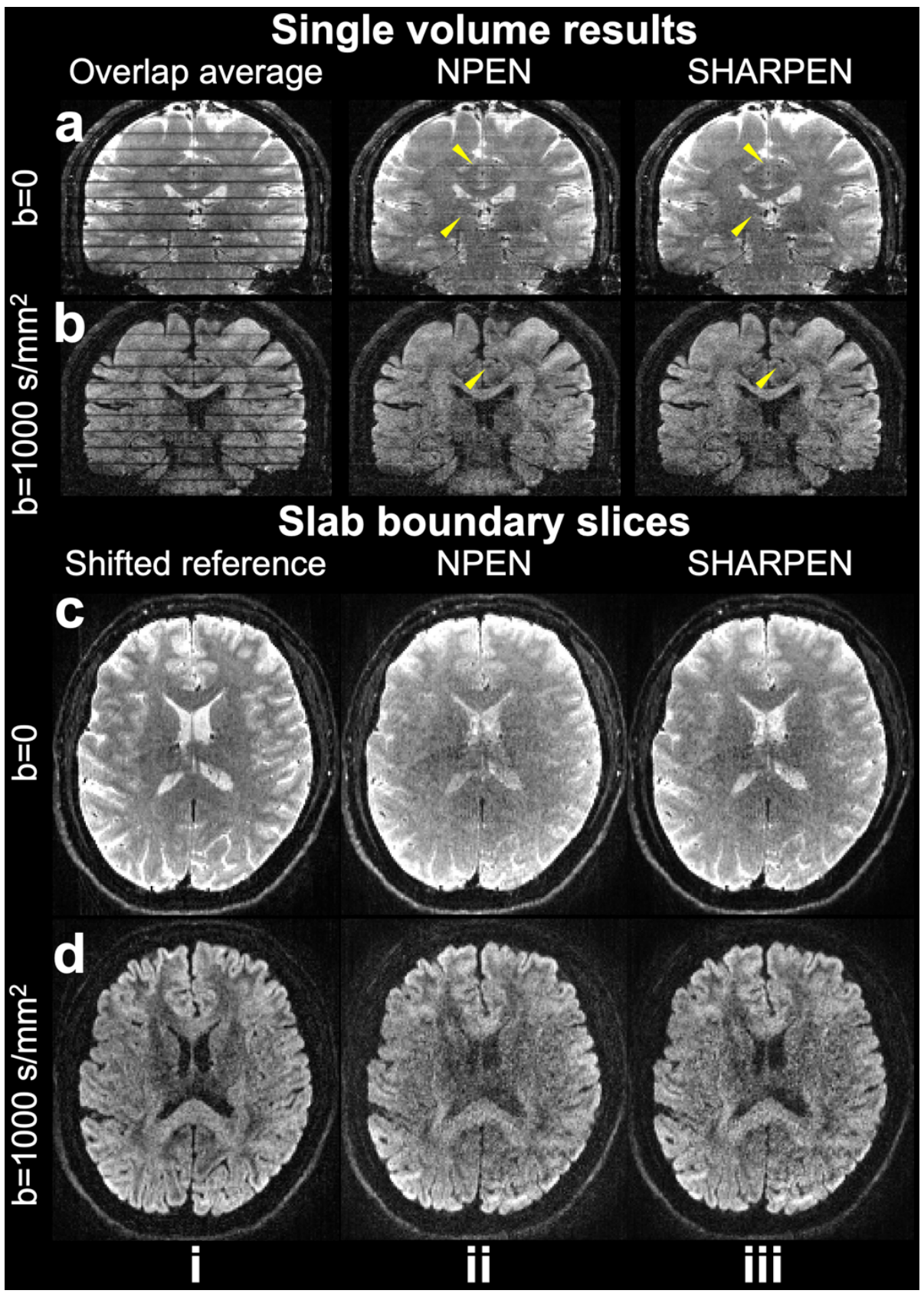

**Figure 6. Single-volume image correction results on prospective 1.1 mm data.** Representative coronal slices from prospective 1.1 mm data are shown for b=0 (a) and b=1000 s/mm$^2$ (b) using overlap averaging (i), NPEN (ii), and SHARPEN (iii). Representative axial slab boundary slices are also shown for b=0 (c) and b=1000 s/mm$^2$ (d), comparing NPEN (i) and SHARPEN (ii) with the corresponding slab-center slices from the shifted acquisition (iii), which serve as a reference. Yellow arrows highlight SHARPEN's improvement over NPEN.

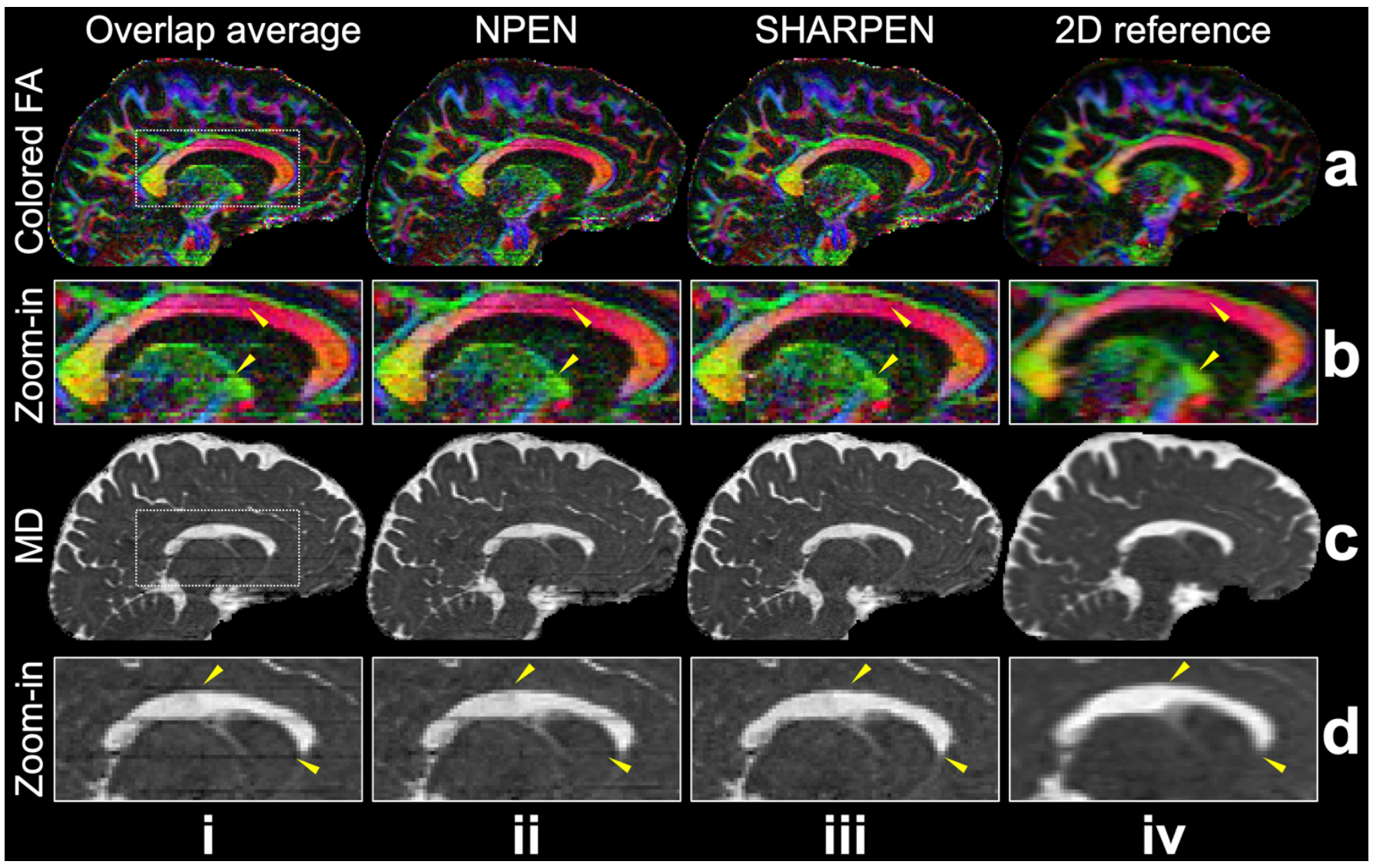

**Figure 7. DTI fitting results on prospective 1.1 mm data.** Representative sagittal slices of V1-modulated fractional anisotropy (colored FA) (a) and mean diffusivity (MD) (c) maps derived from prospective 1.1 mm data are shown for different correction strategies: overlap average (i), NPEN (ii), SHARPEN (iii), and reference 2D datasets (iv). Enlarged regions show image detail, with yellow arrows marking SHARPEN's improvement over other methods.

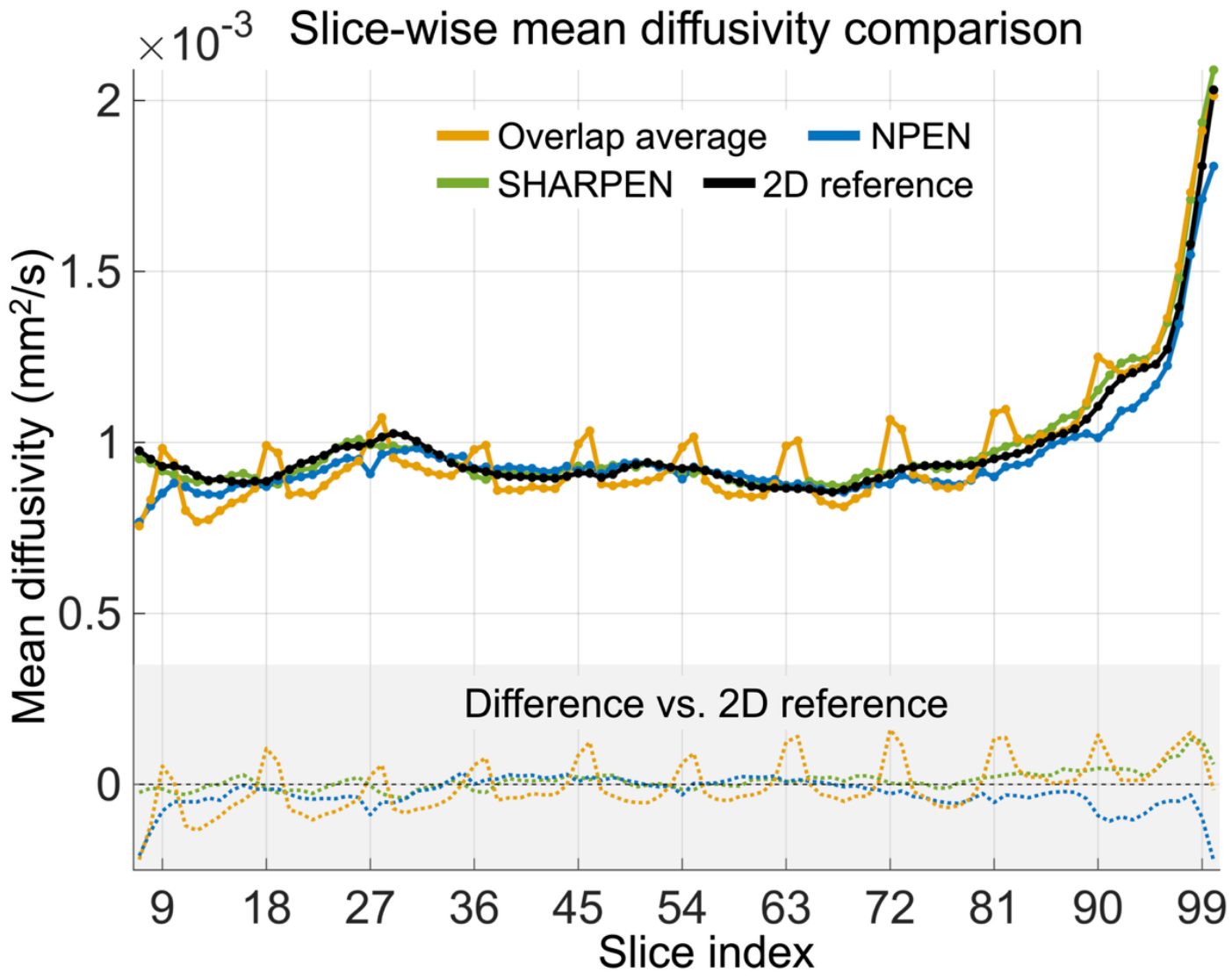

**Figure 8. Slice profile comparison on prospective 1.1 mm data.** Slice-wise mean diffusivity profiles from a representative subject in the prospective 1.1 mm dataset are shown for different slab boundary correction strategies: overlap averaging (yellow), NPEN (blue), SHARPEN (green), and the 2D reference dataset (black). The slice-wise differences relative to the 2D reference are shown in the lower panel (gray shaded region) using dotted lines. Only slices containing valid brain tissue across all acquisitions are displayed.

Slice-wise comparison with the 2D reference dataset further supports the accuracy of SHARPEN-corrected 3D multi-slab data (Fig. 8). To avoid contrast differences arising from mismatched TE/TR between the 2D and 3D acquisitions, slice-wise MD values were compared instead of raw image intensities, providing effective contrast normalization. The overlap average results exhibit pronounced periodic deviations from the 2D reference, reflecting residual slab boundary artifacts, which are partially suppressed by NPEN. However, NPEN introduces a mild bias near the superior brain slices, likely due to its tendency to flatten signal variations across slices, consistent with simulation results (Fig. 5). SHARPEN yields slice profiles that most closely match the 2D reference throughout the brain, demonstrating superior artifact suppression and reconstruction fidelity. Quantitatively, the group-level Pearson correlation coefficients between the 3D multi-slab and 2D reference slice profiles were r=0.9572±0.0231, 0.9732±0.0196, and 0.9904±0.0078 (mean±std) for overlap average, NPEN, and SHARPEN, respectively, across four subjects with prospective 1.1 mm acquisitions. This ordering ($r_{\text{SHARPEN}} > r_{\text{NPEN}} > r_{\text{overlap average}}$) was consistent across all subjects.

SHARPEN demonstrates robustness to inter-volume motion between slab-shifted acquisitions (Fig. 9). As a baseline comparison, we evaluated a conventional direct 'cut-and-combine' approach, which corrects slab boundary artifacts by retaining slab-center slices from each shifted acquisition. When the shifted volume is motion-free, both direct combination and SHARPEN yield high-quality correction results (Fig. 9a, i), and the single-volume SHARPEN correction closely resembles the direct combination result obtained from two shifted volumes. However, when a motion-corrupted shifted volume acquired at the end of the scan and affected by inter-volume motion (4.3 mm translation and ~1° rotation relative to the non-shifted volume) is used, direct combination produces visible anatomical inconsistencies (Fig. 9a, yellow arrows). In contrast, SHARPEN remains robust to this motion, producing corrected images that closely match those obtained using motion-free shifted data (Fig. 9b). This robustness is further reflected in the substantially lower NRMSE observed for SHARPEN compared with direct combination (8.0% vs. 20.9%) when comparing motion-corrupted and motion-free results.

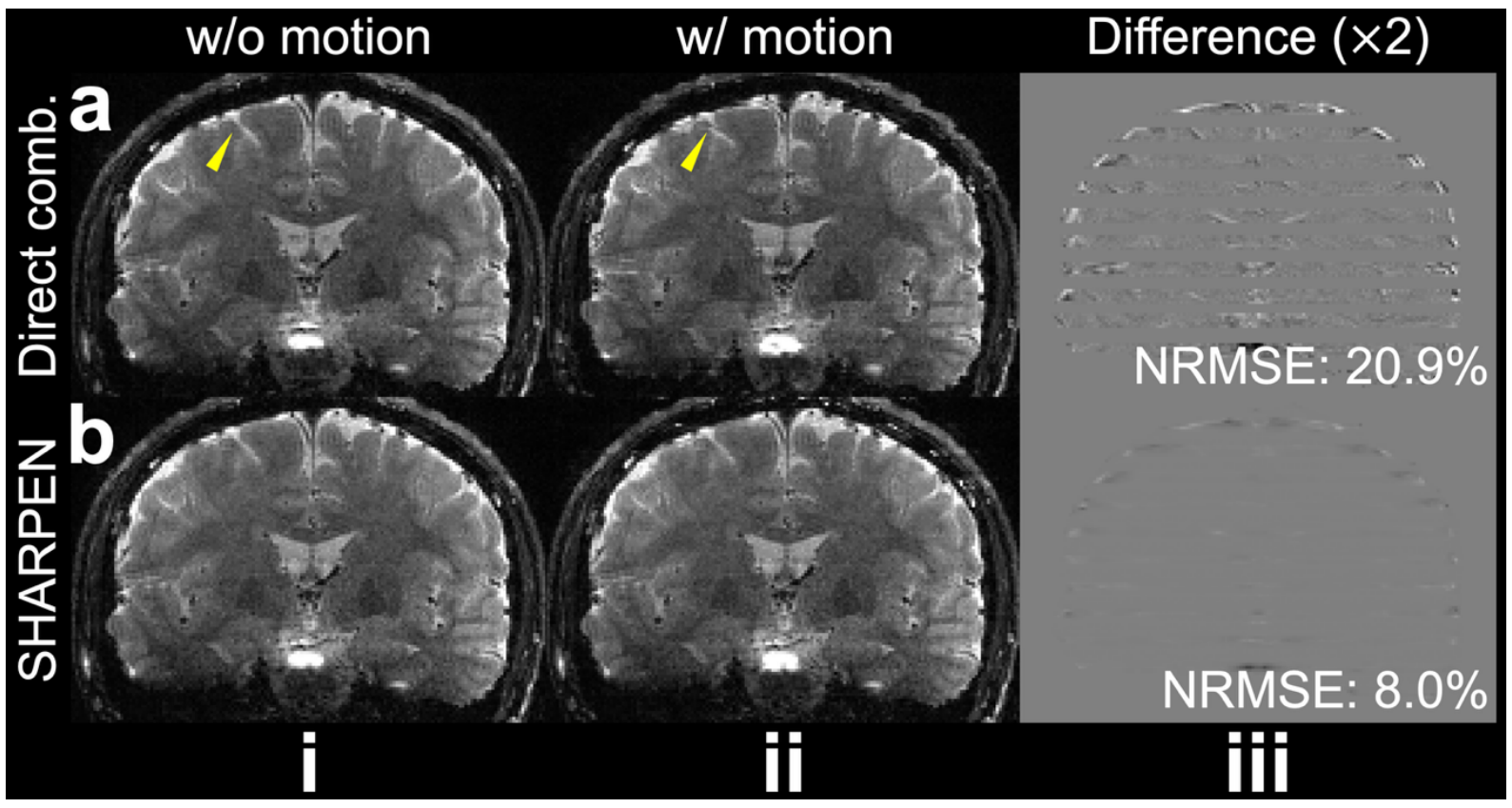

**Figure 9. Inter-volume motion robustness.** Representative coronal slices from a b=0 volume corrected using direct 'cut-and-combine' of two shifted acquisitions (a) and SHARPEN (b) are shown for cases without (i) and with (ii) inter-volume motion. Corresponding difference maps (iii; displayed at 2× intensity scaling) illustrate sensitivity to motion. Normalized root mean squared error (NRMSE), computed within the brain mask, quantifies image similarity between the motion-free and motion-corrupted reconstructions. Yellow arrows highlight anatomical inconsistencies introduced by direct combination when motion is present.

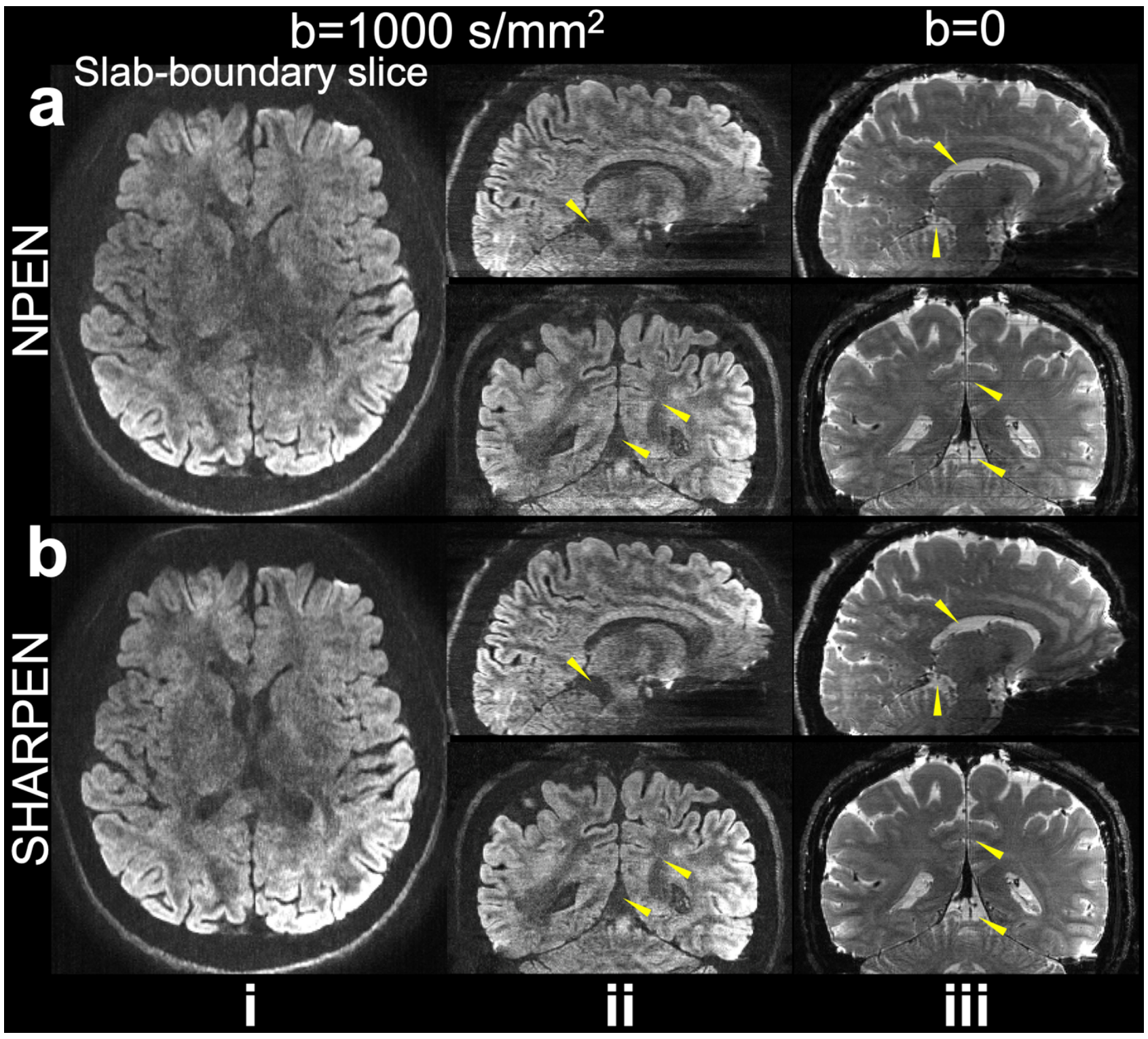

**Figure 10. Single-volume image correction results on prospective 0.7 mm data.** Representative slab boundary axial slices (i), sagittal, and coronal (ii, iii) slices from prospective 0.7 mm data are shown comparing NPEN (a) and SHARPEN (b) correction for b=1000 s/mm$^2$ (i, ii) and b=0 (iii) data. Yellow arrows highlight SHARPEN's improvement over NPEN.

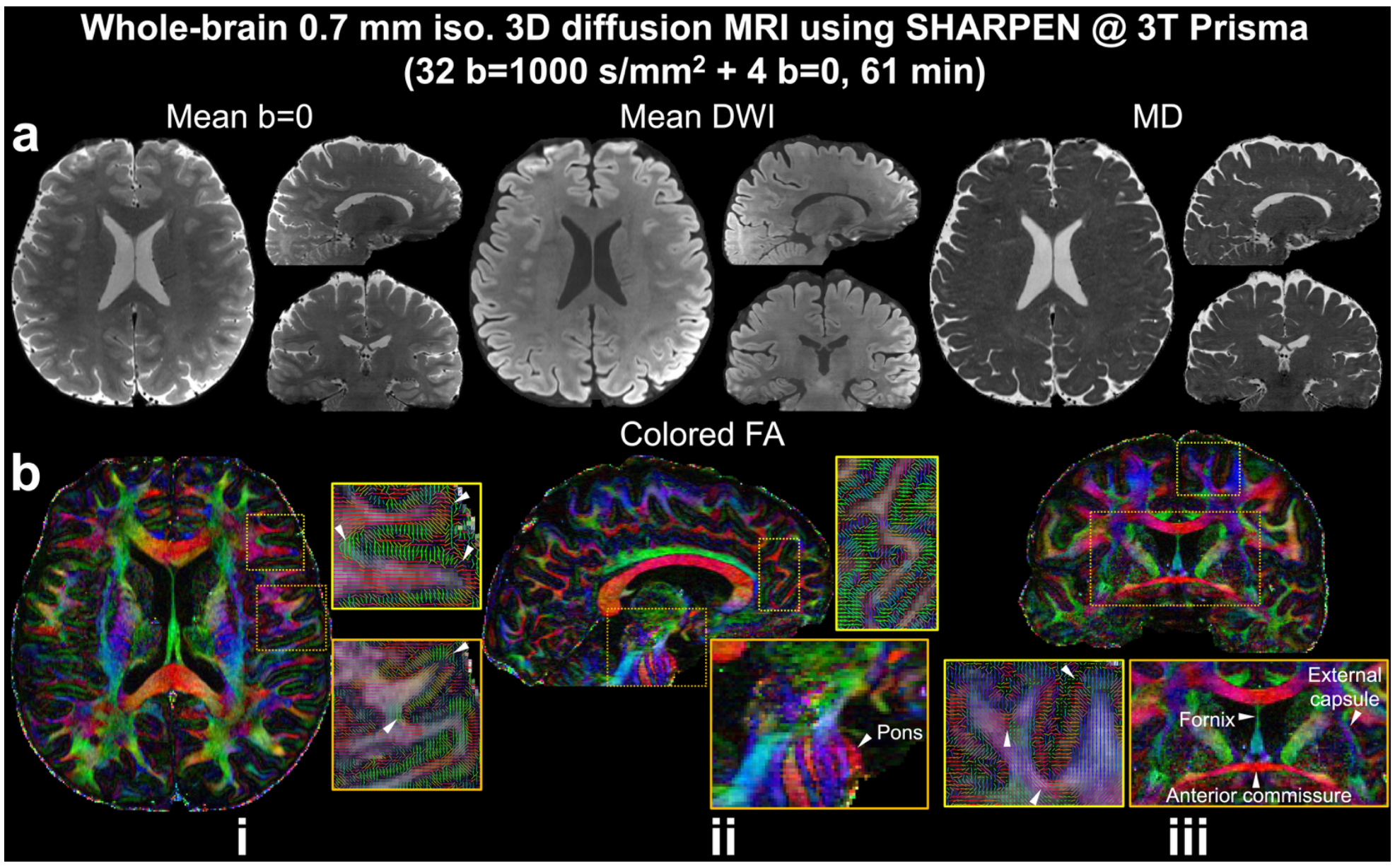

**Figure 11. High-fidelity 0.7 mm dMRI results from SHARPEN.** Mean b=0 (i), mean diffusion-weighted images (DWI) (ii), mean diffusivity (MD) (iii) maps (a), and V1-modulated fractional anisotropy (colored FA) maps (b) from 0.7 mm SHARPEN data demonstrate the high image quality. All axial slices are shown at slab-boundary locations. White arrows highlight fine structures resolved at the ultrahigh resolution.

On the prospectively acquired submillimeter 0.7 mm dMRI data, SHARPEN also delivers high-quality results (Figs. 10, 11). Single-volume reconstructions exhibit adequate SNR and sharp anatomical detail, enabled by the segmented 3D multi-slab acquisition in combination with DnSPIRiT reconstruction (Fig. 10). Compared with NPEN, SHARPEN achieves markedly improved slab boundary correction with reduced noise amplification at slab boundaries (Fig. 10, i) and fewer residual artifacts in coronal and sagittal views (Fig. 10, ii, iii, yellow arrows). These improvements translate into a high-quality submillimeter dataset acquired on a 3T clinical scanner within a one-hour scan time (Fig. 11). The mean b=0, mean DWI, and MD maps exhibit minimal residual artifacts and high overall image quality (Fig. 11a). Color-coded FA maps further delineate fine anatomical structures, including pontine fibers, the fornix, external capsule, and anterior commissure, as well as complex fiber architectures such as fanning at gyral crowns and U-shaped short association fibers (Fig. 11b, white arrows), even in regions with strong B0/B1+ field inhomogeneity (e.g., the pons). Notably, all axial slices shown in Fig. 11 are taken from slab boundary locations, underscoring SHARPEN's ability to produce high-quality results even at slab boundaries. Furthermore, "eddy"-estimated rigid-body motion parameters for this dataset indicate cumulative head translations of up to ~4 mm, including ~2 mm through-slice displacement, and rotations approaching ~1° over the one-hour acquisition. Despite such inter-volume motion, SHARPEN consistently produces high-quality reconstructions across the entire dataset, highlighting its robustness to realistic subject motion during extended submillimeter 3D multi-slab acquisitions.

## 5. Discussion

In this study, we propose SHARPEN, a framework that combines slab-shifted acquisition with self-supervised deep learning to enable accurate correction of slab boundary artifacts in 3D multi-slab dMRI. The inter-volume slab-shifting strategy introduces complementary slab-profile encoding information without increasing scan time, while the self-supervised learning formulation exploits this information to estimate accurate slab profiles without requiring high-quality, artifact-free reference data. The lightweight network can be trained in a subject-specific manner using only a single pair of slab-shifted images from each subject. Validated on both simulated data and prospectively acquired in vivo datasets, SHARPEN achieves higher correction accuracy and more than an order-of-magnitude reduction in processing time, including network training, compared with NPEN. In addition, SHARPEN demonstrates robustness to inter-volume motion during acquisition. Its ability to produce high-quality 0.7 mm isotropic dMRI on a clinical 3T scanner within a one-hour session underscores its potential to enable high-fidelity submillimeter dMRI for neuroscience and clinical research.

The inter-volume slab-shifting strategy efficiently captures complementary slab profile encoding at both slab centers and boundaries. Unlike prior intra-volume slab-shifting approaches [9-11], which require multiple shifts per volume and retain only slab-center slices, our method applies different slab shifts across diffusion directions, enabling full use of all acquired slices without increasing scan time. Within SHARPEN, this strategy underpins what we refer to as slab-shifted harmonization: information from different slab shifts is jointly leveraged during acquisition and reconstruction to harmonize slice quality across the volume. Slab-shifted harmonization provides two key benefits. First, complementary encoding across shifts enables accurate self-supervised slab profile estimation without requiring high-quality reference data. Second, it reduces slice-wise error variance in downstream diffusion analyses: even at slab boundaries in one shift group, half of the corresponding slices originate from slab centers in the other group, effectively improving estimation robustness. This effect is quantitatively demonstrated in HCP simulations (Fig. 5), where SHARPEN achieves both lower slice-wise errors and reduced error variance. While slices that consistently fall at slab centers retain inherently higher SNR than boundary slices, slab-shifted harmonization substantially elevates boundary-slice SNR and mitigates the pronounced degradation typically observed at slab boundaries. The advantage is particularly pronounced in low-SNR settings (e.g., submillimeter or high-b-value imaging), where slab boundary slices approach the noise floor. Our submillimeter in-vivo results (Fig. 11) further validate this benefit, demonstrating high-quality reconstructions even at slab boundaries.

Slab-shifted acquisition in SHARPEN redistributes diffusion measurements over an expanded field-of-view along the slice direction. As a result, the extreme superior and inferior half-slabs are sampled with half of the diffusion directions, while the total number of acquired measurements remains unchanged. This represents an explicit trade-off of the proposed acquisition strategy. In practice, analyses can be focused on slices with full directional coverage, where SHARPEN provides substantially improved image fidelity and reduced slab-boundary artifacts. Excluding a small number of peripheral slices is an acceptable cost of improved data quality within the central brain regions. For completeness, full-FOV results are shown in Supplementary Fig. 4, demonstrating that SHARPEN maintains reasonable image quality at the inferior edge despite reduced directional sampling and strong B0/B1+ inhomogeneity, although data quality is expectedly lower in these regions. Looking ahead, integrating SHARPEN with advanced acquisition strategies that shorten TR, such as self-navigated 3D multi-slab dMRI [22] and simultaneous multi-slab dMRI [30, 31], may further extend high-quality coverage without increasing scan times.

Our self-supervised learning formulation hinges on two key principles: consistency across slab-shifted acquisitions and prior physical knowledge of slab profiles and diffusion images. Consistency across shifts ($\mathcal{L}_{\mathrm{shift}}$) is enforced by assuming that mean corrected images from different shift groups exhibit similar trace-weighted contrast and that slab excitation profiles are invariant up to spatial displacement, as tissue composition and field inhomogeneities vary slowly in space. This inter-volume redundancy enables accurate profile estimation without requiring external reference data. Additional constraints encode prior knowledge of the acquisition and image formation process. Specifically, the slab-center profile is constrained to be flat via $\mathcal{L}_{\mathrm{profile}}$, an assumption well supported by modern RF pulse design (e.g., SLR pulses used in this study). In-plane smoothness of the slab profile is enforced through $\mathcal{L}_{\mathrm{smooth}}$, as in prior profile encoding approaches, to improve conditioning of the inverse problem [12–14]. Finally, slice-to-slice intensity variation is penalized by $\mathcal{L}_{\mathrm{slice}}$ only at slab boundaries to avoid over-regularization. These complementary constraints enable robust slab profile estimation in a fully self-supervised manner, explaining SHARPEN's strong performance across resolutions, SNR regimes, and motion conditions.

The adoption of a lightweight CNN promotes SHARPEN's data and computational efficiency. Theoretically, SHARPEN could be formulated as a nonlinear optimization problem and solved using iterative methods such as Gauss-Newton, similar to NPEN [13]. We instead adopt a neural network formulation to leverage its superior ability to model nonlinear mappings and the availability of efficient, GPU-accelerated training frameworks (e.g., PyTorch [26]). A compact CNN is sufficient because the slab profile to be estimated is smooth and low in complexity. With ~0.1 million parameters, the network can be trained in a subject-specific, self-supervised manner using only a single pair of slab-shifted volumes. This is because training is voxel-wise, and each voxel serves as an independent sample. Even the 1.25 mm HCP data provide over one million effective samples, yielding a favorable sample-to-parameter ratio (>10) that enables fast and stable training (Supplementary Fig. 2). As a result, SHARPEN achieves more than an order-of-magnitude speedup compared with NPEN, even when NPEN is parallelized across many CPU cores, substantially improving its practical feasibility.

SHARPEN demonstrates robust performance in the presence of inter-volume motion from two complementary perspectives. First, it remains effective when motion occurs between slab-shifted acquisitions, producing accurate correction even with a motion-corrupted shifted volume (Fig. 9). Second, SHARPEN is robust when a single slab profile estimated from mean DWI is applied across all DWI volumes, even for an extended one-hour acquisition with moderate subject motion (Supplementary Fig. 5), where "eddy"-estimated parameters indicate head translations of up to 4 mm (including 2 mm through-slice displacement) and rotations approaching 1°. This robustness is enabled by the smooth spatial variation of slab excitation, saturation, and B0/B1+ field inhomogeneity [12], which SHARPEN explicitly exploits during profile estimation. In particular, the shift-consistency ($\mathcal{L}_{\mathrm{shift}}$) assumption that slab profiles remain stable over a half-slab displacement holds under realistic inter-volume motion and supports reliable correction across time. We note that SHARPEN does not explicitly address intra-volume rigid-body motion that leads to spatial misalignment between shots, which remains an intrinsic challenge for 3D multi-slab dMRI, despite the use of navigator-based phase correction. Integrating SHARPEN with advanced intra-volume motion correction strategies for 3D EPI [32] represents a promising direction for future work.

## 6. Conclusion

We have presented SHARPEN, a slab boundary artifact correction framework for 3D multi-slab diffusion MRI that integrates slab-shifted acquisition with self-supervised deep learning. By exploiting consistency across slab shifts and known physical properties of slab excitation profiles and diffusion images, SHARPEN enables accurate slab boundary correction without requiring high-quality reference training data. The use of a lightweight CNN allows subject-specific training, while an efficient GPU-based implementation provides fast and accurate correction. As a result, SHARPEN yields slice-wise quantitative profiles along the slice direction that closely match those obtained from reference 2D acquisitions. SHARPEN also demonstrates robustness to inter-volume motion and performs reliably across both simulated and prospectively acquired high-resolution in vivo datasets. Finally, the ability to produce high-quality submillimeter dMRI on a clinical scanner highlights SHARPEN's potential to advance imaging neuroscience through more detailed characterization of human brain microstructure.

**Acknowledgments**

K.L.M. and W.W. contributed equally to this work. MRI sequences in this work were implemented using the scanner-agnostic open-source platform Pulseq. We used diffusion and T1w MRI data from the Human Connectome Project, WU-Minn-Ox Consortium (Principal Investigators: David Van Essen and Kamil Ugurbil; U54-MH091657) funded by the 16 NIH Institutes and Centers that support the NIH Blueprint for Neuroscience Research, and by the McDonnell Center for Systems Neuroscience at Washington University. W.W. is supported by Wellcome Trust (WT327832/Z/25/Z) and was supported by the Royal Academy of Engineering (RF\201819\18\92). K.L.M. is supported by the Wellcome Trust (WT224573/Z/21/Z and WT215573/Z/19/Z). This study is supported by the NIHR Oxford Health Biomedical Research Centre (NIHR203316). The views expressed are those of the author(s) and not necessarily those of the NIHR or the Department of Health and Social Care.

**Supplementary Information**

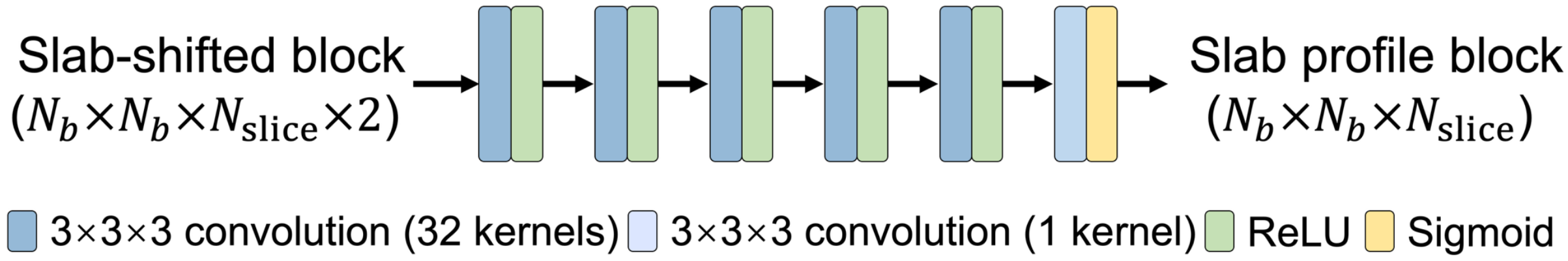

**Supplementary Figure 1. SHARPEN network architecture.** SHARPEN employs a lightweight 3D convolutional neural network (CNN) to estimate the slab excitation profile from a concatenated slab-shifted, slice position-aligned input block. The block size $N_b$ controls the spatial extent of the input, and block-wise processing is used to reduce GPU memory requirements. A final sigmoid function constrains the estimated slab profile to the physically meaningful range (0, 1).

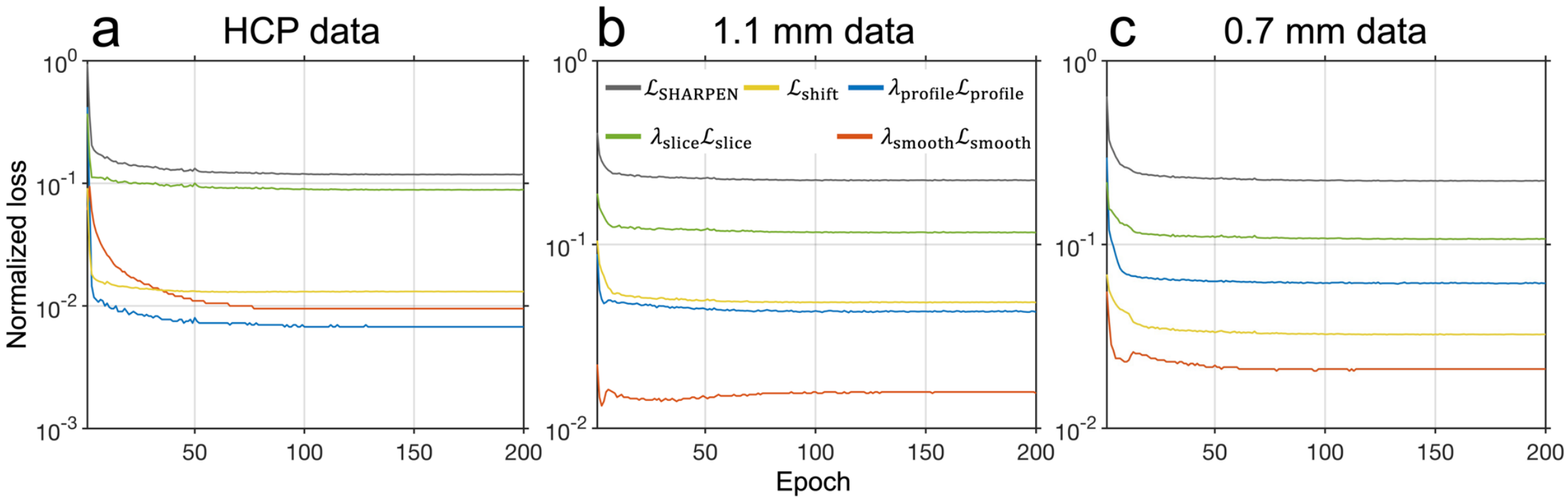

**Supplementary Figure 2. SHARPEN training loss convergence.** Normalized training losses for SHARPEN are shown for one representative subject from HCP simulation (a), prospectively acquired 1.1 mm (b), and 0.7 mm data (c). The curves show the weighted total loss $\mathcal{L}_{\text{SHARPEN}}$ (gray) and the corresponding weighted loss components $\mathcal{L}_{\text{shift}}$ (yellow), $\lambda_{\text{profile}}\mathcal{L}_{\text{profile}}$ (blue), $\lambda_{\text{slice}}\mathcal{L}_{\text{slice}}$ (green), and $\lambda_{\text{smooth}}\mathcal{L}_{\text{smooth}}$ (red), illustrating stable and consistent convergence across datasets.

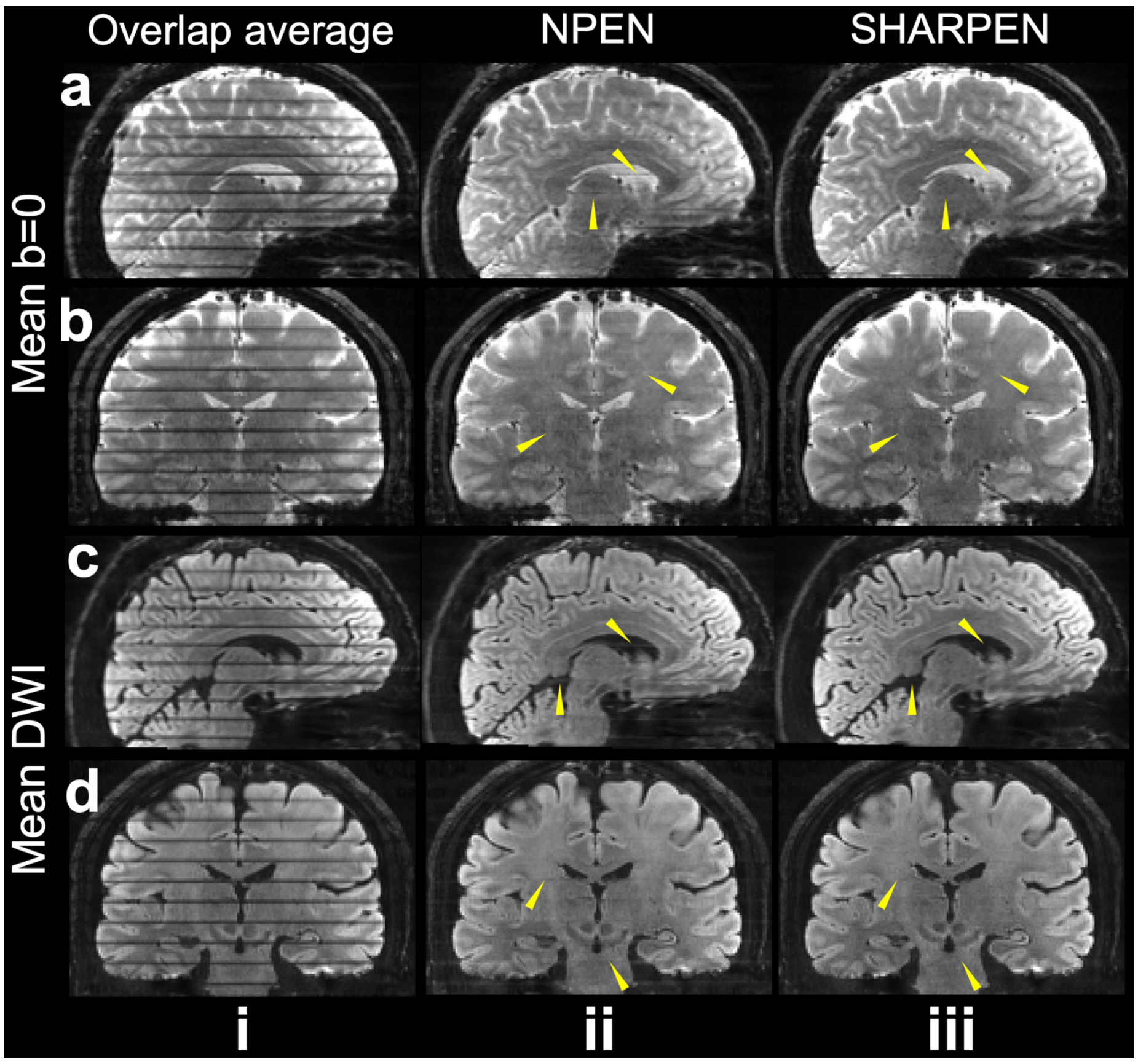

**Supplementary Figure 3. Corrected mean images on prospective 1.1 mm data.** Representative sagittal (a, c) and coronal (b, d) slices from prospective 1.1 mm data are shown for mean b=0 (a, b) and mean diffusion-weighted images (DWI) (c, d) using overlap averaging (i), NPEN (ii), and SHARPEN (iii) from the prospective 1.1 mm data. Yellow arrows highlight SHARPEN's improvement over NPEN.

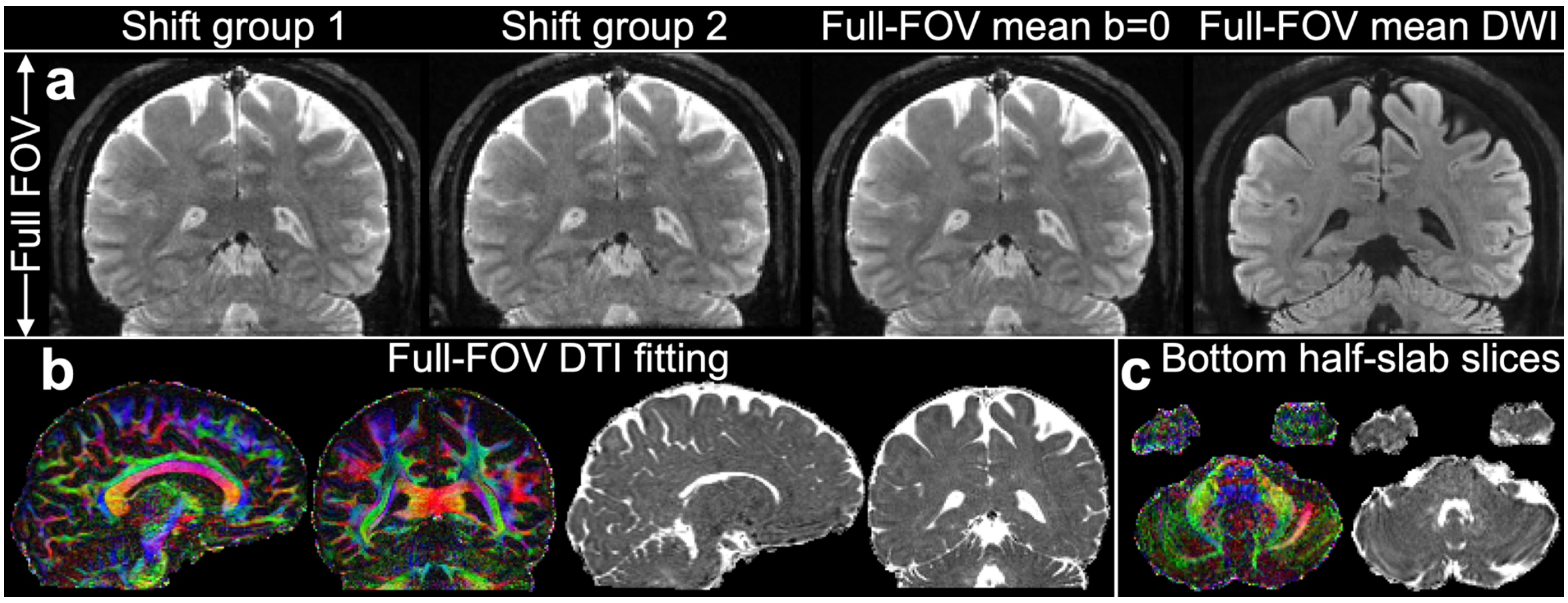

**Supplementary Figure 4. SHARPEN full-FOV results.** Representative coronal b=0 images from prospective 1.1 mm data illustrate the slab shift between two SHARPEN shift groups and the resulting full-field-of-view (FOV) mean b=0 and mean diffusion-weighted images (DWI) (a). At the superior and inferior FOV extremes, only one shift group contributes, resulting in two half-slabs with half the diffusion directions. Full-FOV diffusion tensor imaging (DTI) results including color-coded fractional anisotropy and mean diffusivity maps (b) and an axial slice from the inferior half-slab (c) demonstrate preserved image quality at the volume edge.

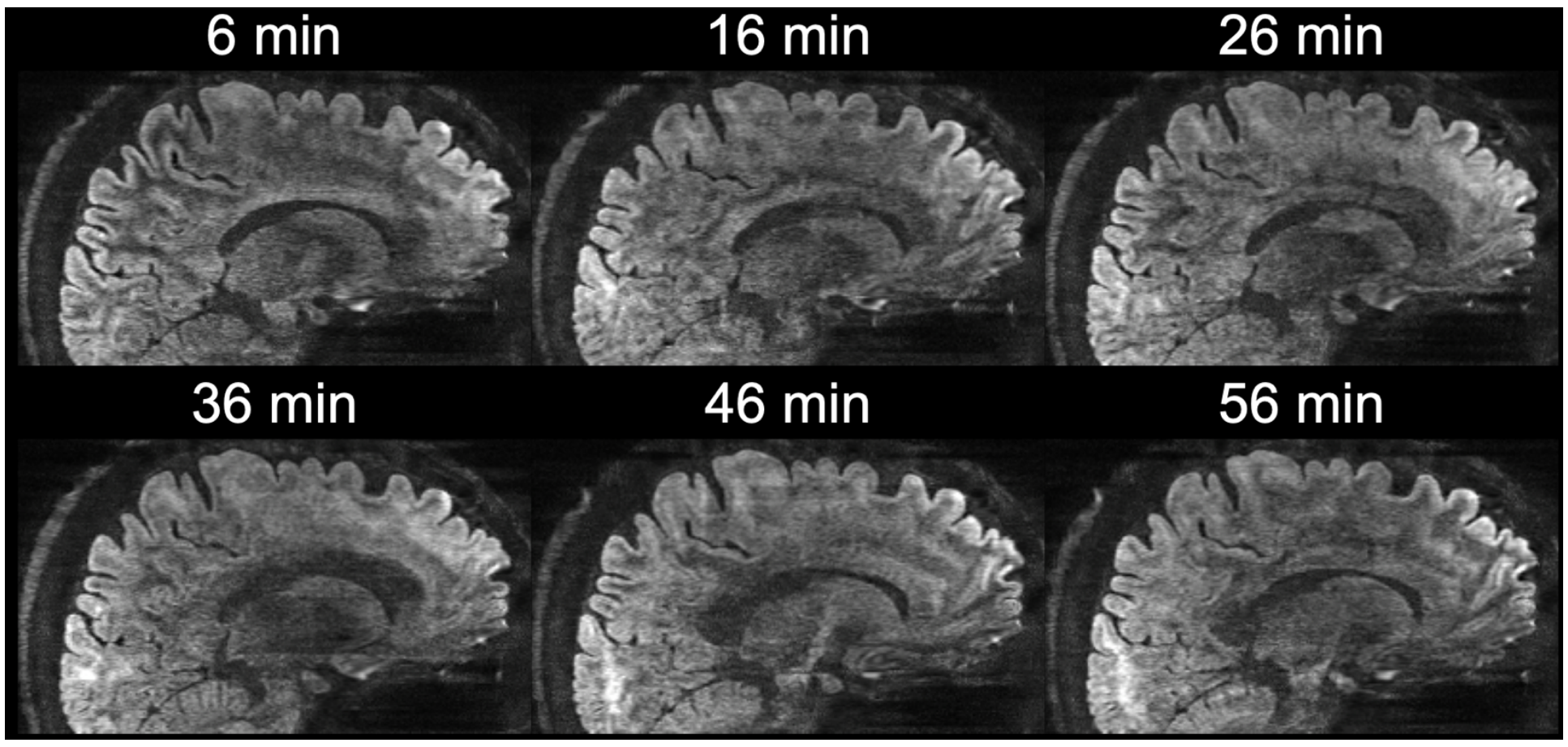

**Supplementary Figure 5. SHARPEN robustness over a one-hour acquisition.** Representative sagittal diffusion-weighted images reconstructed with SHARPEN from prospective 0.7 mm data are shown at different time points throughout the one-hour acquisition. Despite moderate inter-volume motion over time, image quality remains consistent, demonstrating SHARPEN's robustness when applying a single slab profile across extended high-resolution scans.